\documentclass[sigconf]{acmart}

\usepackage{booktabs} 

\usepackage{multirow}
\usepackage{wrapfig}
\usepackage{algorithmic}
\usepackage{amsfonts}
\usepackage{amsmath}
\usepackage{siunitx}

\usepackage{soul}
\usepackage{caption}
\usepackage{subcaption}

\newcommand{\ourmethod}{Gravo MG}

\usepackage[ruled]{algorithm2e} 

\SetAlFnt{\small}
\SetAlCapFnt{\small}
\SetAlCapNameFnt{\small}
\SetAlCapHSkip{0pt}
\copyrightyear{2023}
\acmYear{2023}
\setcopyright{rightsretained}
\acmConference[SIGGRAPH '23 Conference Proceedings]{Special Interest Group on Computer Graphics and Interactive Techniques Conference Conference Proceedings}{August    6--10, 2023}{Los Angeles, CA, USA}
\acmBooktitle{Special Interest Group on Computer Graphics and Interactive Techniques Conference Conference Proceedings (SIGGRAPH '23 Conference Proceedings), August    6--10, 2023, Los Angeles, CA, USA}
\acmDOI{10.1145/3588432.3591502}
\acmISBN{979-8-4007-0159-7/23/08}
\begin{document}

\title{A Fast Geometric Multigrid Method for Curved Surfaces}

\author{Ruben Wiersma}
\orcid{0000-0001-7900-7253}
\email{r.t.wiersma@tudelft.nl}
\authornote{Both authors contributed equally to this research.}
\affiliation{%
 \institution{Delft University of Technology}
 \country{The Netherlands}
 }
\author{Ahmad Nasikun}
\orcid{0000-0001-5311-4456}
\email{ahmad.nasikun@ugm.ac.id}
\authornotemark[1]
\affiliation{%
 \institution{Universitas Gadjah Mada}
 \institution{Delft University of Technology}
 \country{Indonesia, The Netherlands}
 }
\author{Elmar Eisemann}
\orcid{0000-0003-4153-065X}
\email{e.eisemann@tudelft.nl}
\affiliation{%
 \institution{Delft University of Technology}
 \country{The Netherlands}
 }
\author{Klaus Hildebrandt}
\orcid{0000-0002-9196-3923}
\email{k.a.hildebrandt@tudelft.nl}
\affiliation{%
 \institution{Delft University of Technology}
 \country{The Netherlands}
 }
\renewcommand\shortauthors{R. Wiersma, A. Nasikun, E. Eisemann, \& K. Hildebrandt}
%
%


%
%


\begin{teaserfigure}
\includegraphics[width=\textwidth]{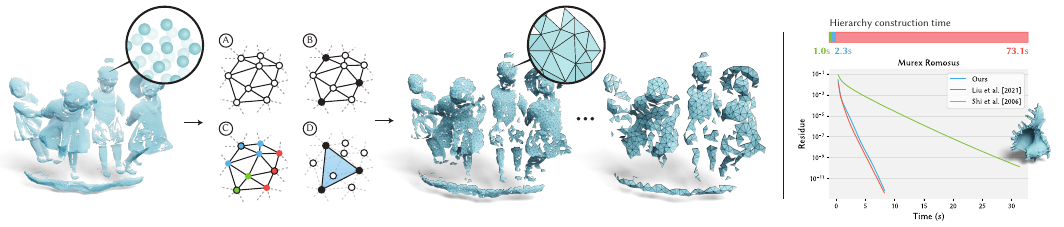}
\caption{Illustration of our hierarchy construction for one level. Starting from a mesh or point cloud, A) we 
construct a neighbor graph on the surface. B) We then sample a spatially uniform set of nodes, C) compute the graph Voronoi diagram of the samples, and D) project unsampled points onto triangles formed by edges between Voronoi neighbors. This is repeated for every level. Right: Comparison of run time for solving a Laplace system on a triangle mesh. Our hierarchy construction is fast, while achieving similar solver performance to the state-of-the-art.}
\label{fig:teaser}
\end{teaserfigure}


\begin{abstract}
We introduce a geometric multigrid method for solving linear systems arising from variational problems on surfaces in geometry processing, \ourmethod. Our scheme uses point clouds as a reduced representation of the levels of the multigrid hierarchy to achieve a fast hierarchy construction and to extend the applicability of the method from triangle meshes to other surface representations like point clouds, nonmanifold meshes, and polygonal meshes. To build the prolongation operators, we associate each point of the hierarchy to a triangle constructed from points in the next coarser level. We obtain well-shaped candidate triangles by computing graph Voronoi diagrams centered around the coarse points and determining neighboring Voronoi cells. Our selection of triangles ensures that the connections of each point to points at adjacent coarser and finer levels are balanced in the tangential directions. As a result, we obtain sparse prolongation matrices with three entries per row and fast convergence of the solver. Code is available at \url{https://graphics.tudelft.nl/gravo_mg}.
\end{abstract}

\maketitle

\section{Introduction}
Many geometry processing methods are based on variational problems and partial differential equations on curved surfaces. The discretization of these problems leads to sparse linear systems to be solved. One class of efficient solvers are Geometric Multigrid (GMG) methods, which use iterative solvers on a hierarchy of grids. They are more efficient than alternatives, such as sparse direct solvers, in many application scenarios \cite{Liu2021}. 
While geometric multigrid solvers are well-studied for regular grids in Euclidean domains, the construction of effective geometric multigrid hierarchies remains challenging for irregular meshes on curved domains. 

We distinguish two approaches to the design of GMG methods on curved surfaces. 
The first approach is to construct a hierarchy of meshes by mesh coarsening and then mapping between the meshes. This approach obtains efficient prolongation operators that lead to fast convergence. A recent example is the intrinsic multigrid scheme by \citet{Liu2021}. The downside of this approach is a costly hierarchy construction. 
The second approach is to represent levels by graphs constructed by coarsening the edge graph of the input mesh \cite{Shi2006}. This approach results in a fast construction but slower convergence. 

We propose a new GMG method combining the strengths of both approaches. On the one hand, we use point clouds and neighbor graphs to represent levels, enabling a fast hierarchy construction. On the other hand, we use geometric operations to create local triangulations when constructing the prolongation operators for fast convergence.  
Our method solves linear systems as fast as the scheme of \citet{Liu2021}, while reducing hierarchy-construction time by more than an order of magnitude.
Moreover, our method is more generally applicable as it can be used not only for manifold triangular meshes but also for other discrete surface representations such as point clouds, non-manifold meshes, and polygonal meshes. Thus, we can solve systems set up with discrete differential operators for these representations, which were developed in recent years \cite{Liang2013,Sharp2020,Alexa2011}. Our hierarchy construction is more expensive compared to \citet{Shi2006}. Yet, the solving time is most often reduced more than the increase in hierarchy construction. This benefit increases for applications where multiple systems need to be solved.

The technical novelty of our method lies in a geometric multigrid method that is point-based, while still incorporating the geometry of the underlying surface.
Our guiding idea is to construct intrinsic Delaunay triangulations on points sampled from the surface. Every other point can then be mapped from- and to the sampled points using barycentric coordinates in the intrinsic triangles. 
To get a fast and practical approach, we transfer this idea to a point-cloud setting. 
For every level in the hierarchy, we start by sampling points from the previous level using a fast uniform sampling strategy. Next, we compute graph Voronoi diagrams on the finer level using the sampled points as seeds and construct a neighborhood graph based on Voronoi cell adjacencies. Mimicking Delaunay triangulations, we construct triangles from the edges of the Voronoi adjacency graph.
Each point of the finer level is projected to its closest triangle and barycentric coordinates are used for prolongation. 
This construction leads to sparse prolongation matrices with at most three entries per row and hence to fast prolongations and restrictions. The use of graph Voronoi cells ensures that the prolongation matrix and its transpose (the restriction matrix) contain entries corresponding to neighbors that are well-distributed over the tangential directions. We name our hierarchy construction \ourmethod, for graph Voronoi multigrid.

We evaluate \ourmethod~in ablations and comparisons to \cite{Liu2021}, \cite{Shi2006}, and algebraic multigrid methods. Furthermore, we demonstrate the benefits of our scheme over sparse direct solvers in application scenarios. 

\section{Related Work}

\paragraph{Geometric multigrid}
Multigrid methods \cite{Bramble1993} are among the most efficient iterative methods for solving linear systems. We call them  \textit{geometric} multigrid methods if the hierarchy construction is exclusively on the domain and no information is used about the system to be solved.  
GMG methods on regular grids are well studied \cite{Hackbusch1985} and used in graphics, \emph{e.g.}, for fluid simulation \cite{McAdams2010,Dick2016}, image processing \cite{Perez2003,Krishnan2011}, and surface reconstruction \cite{Kazhdan2006,Kazhdan2019}. 
GMG methods for irregular grids on Euclidean domains are used for the simulation of cloth (2D) \cite{Jeon2013,Wang2018} and elastic objects (3D) \cite{Georgii2006,Otaduy2007}.

In this work, we consider GMG methods for curved surfaces. Since the domain is no longer a Euclidean space but a curved manifold, methods from the Euclidean setting do not transfer directly and new methods are needed. Existing GMG methods on surfaces focus on triangle mesh representations of discrete surfaces.   
If the mesh is already equipped with a hierarchy, for example from a subdivision method, it can be used directly for a multigrid method \cite{Green2002}.
However, usually, only a fine-scale mesh is given and a hierarchy must be built. 
Based on earlier work on multiresolution representations of triangle meshes, \cite{Hoppe1996,Kobbelt1998}, edge collapses are used to create multigrid hierarchies in \cite{Ray2003,Aksoylu2005,Ni2004}. 
The prolongation operators are defined by weighted averaging with the one-ring neighbors. 
There are two approaches to guaranteeing that each vertex in a finer level has at least one neighbour in the coarser level: either the coarsening process is restricted to only collapsing edges, so that a maximal independent set of vertices (MIS) is removed \cite{Ni2004,Aksoylu2005}, or the edge collapses are restricted so that a MIS is preserved \cite{Aksoylu2005}. 
\citet{Liu2021} introduce an intrinsic multigrid scheme that uses edge coarsening to create the meshes for the different levels and maintains bijective mappings between the meshes on consecutive levels. 
The map between two meshes is used to define the prolongations. The map assigns to each vertex of the finer mesh a point in a triangle of the coarser mesh and linear interpolation in the triangle is used for prolongation. The resulting prolongation matrix has at most three entries per row. 
An alternative to mesh coarsening is to use graph coarsening for hierarchy construction \cite{Shi2006,Shi2009}.
A multigrid scheme for the computation of Laplace--Beltrami eigenpairs on surfaces is introduced in \cite{Nasikun2022}. 
The hierarchy used for the eigenproblem, however, is much coarser than the hierarchies used for solving linear systems: only two or three levels are used.
A multigrid solver for the computation of harmonic foliations on surfaces is introduced in~\cite{Wang2022}.

\paragraph{Algebraic multigrid}
Algebraic multigrid (AMG) methods \cite{Brandt1986,Stuben2001} are an alternative to GMG. They use the matrix of the linear system to be solved to build the hierarchy instead of using the domain. This has the advantage that AMG can be used for problems coming from arbitrary domains. 
Nevertheless, AMG methods need to rebuild the hierarchy when the system matrix changes, whereas GMG methods only need to rebuild the hierarchy when the domain changes. 
An efficient multigrid preconditioner specifically for Laplace systems on images and meshes was introduced in \cite{Krishnan2013}. Although fast, it has the disadvantage of requiring the Laplace matrices to have only non-positive off-diagonal entries, which is often not satisfied by mesh Laplacians, such as the cotangent-Laplacian \cite{Pinkall1993}. 

\paragraph{Direct solvers}
Sparse direct solvers \cite{Davis2016} are reliable, accurate, and commonly used for Laplace systems in geometry processing.  
Once a factorization of a matrix is computed, these solvers can solve multiple systems with the same matrix but different right-hand sides. 
In special cases, such as low-rank changes of the matrix, the factorization can be updated efficiently \cite{Chen2008,Herholz2018,Herholz2020}. However, substantial changes require a new factorization.    
A disadvantage of these solvers is that they do not scale well neither in terms of memory requirements nor computation time.
In Section~\ref{sect.experiments}, we compare the performance of our method to direct solvers in different scenarios. 

\begin{algorithm}[t]
\DontPrintSemicolon
\LinesNumbered
\KwIn{Matrix $A\in\mathbb{R}^{n\times n}$, initial vector $x\in\mathbb{R}^{n}$, right-hand side $b\in\mathbb{R}^{n}$, error tolerance $\varepsilon$, number of levels $\lambda$, numbers of pre/post-relaxations steps $\nu_{pre},\nu_{post}$}
\KwOut{Solution $x\in\mathbb{R}^{n}$ to the linear system $Ax=b$}
\SetKwFunction{FMain}{Multigrid}
\SetKwProg{Fn}{Function}{:}{}
\Fn{\FMain{$A, x, b, \varepsilon, \lambda, \nu_{pre},\nu_{post}$}}{
	Build prolongation matrices $P_1,P_2,P_3...,P_{\lambda}$\;
	$A_1\gets A$\;
	\For{$l \gets 2$ to $\lambda$}{
		 $A_{l}\gets(P_{l-1})^{\top} A_{l-1} P_{l-1}$\hfill // Build level $l$ matrix\;
	}
    \Repeat{$\left\Vert Ax-b\right\Vert \leq \varepsilon \left\Vert b\right\Vert$ \hfill //Convergence test}{
		$x\gets$\textbf{MGI}($x,b,1$)\;
	}
	\Return $x$\;
}
\textbf{End Function}
\caption{Multigrid solver}
\label{alg.MGS}
\end{algorithm}

\begin{algorithm}[t]
\DontPrintSemicolon
\LinesNumbered
\KwIn{Current iterate $x\in\mathbb{R}^{n_l}$, right-hand side $b\in\mathbb{R}^{n_l}$, level $l$}
\KwOut{New iterate $x\in\mathbb{R}^{n_l}$}
\SetKwFunction{FMain}{MGI}
\SetKwProg{Fn}{Function}{:}{}
\Fn{\FMain{$x,b,l$}}{
	\eIf{$l<\lambda$}{
     $y\gets$\textbf{Relax}($A_l,x,b,\nu_{pre}$)\hfill //Pre-relaxation\;	
     $u\gets$\textbf{MGI}($0,P_l^{\top}(b-A_l y),l+1$)\hfill //Recursive call\;
	$x\gets$\textbf{Relax}($A_l,x + P_l u,b,\nu_{post}$)\hfill //Post-relaxation\;  
	}{
		Solve $A_{\lambda} x = b$ using a direct solver\;
	}
	\Return $x$\;
}
\textbf{End Function}
\caption{Multigrid Iteration ($V$-cycle)}
\label{alg.MGI}
\end{algorithm}

\section{Background: Multigrid Solver}
Multigrid solvers use a hierarchy of grids to solve systems of equations. 
Iterative solvers converge at different speeds for different scales, depending on the resolution of the grid on which they operate.
Thus, by performing iterations on different grids, a multigrid scheme extends the range in which the solver converges particularly fast. Here, we describe a multigrid solver, which will later be used to evaluate our proposed hierarchy and prolongation operators.

We consider the multigrid solver in Algorithm~\ref{alg.MGS}. To solve an $n$-dimensional linear system $Ax=b$ for $x$, it operates on a multigrid hierarchy with $\lambda$ levels, where level 1 is the finest and level $\lambda$ is the coarsest level. A function on the $l^{th}$ grid is represented by a vector in $\mathbb{R}^{n_l}$, where the grid has $n_l$ degrees of freedom. The mappings between the grids are realized by prolongation and restriction matrices. The prolongation matrices $P_l\in\mathbb{R}^{n_{l}\times n_{l+1}}$ map from level $l+1$ to level $l$. We use the transposed matrices of the prolongation matrices $P_l^{\top}$ as restriction matrices. The advantage is that a symmetric matrix $A$ implies that the linear systems in the coarse grid correction, which involve the restricted matrices $P_l^{\top}A_l P_l$, are also symmetric.

The multigrid solver first builds the prolongation matrices. We keep this step abstract at this point but discuss it in detail in the following section. 
In the next step, lines 3-6, the restricted matrices for all levels are constructed. 
After the precomputation, multigrid iterations are executed until convergence of the solution. The multigrid iterations traverse the hierarchy from fine to coarse and back. This process is called a $V$-cycle and is simple but effective. Alternatively, instead of directly going up to the coarsest grid, one could first go back to finer grids. 
Such strategies can help to counteract error accumulation when several levels are traversed and thereby reduce the required number of multigrid iterations. On the other hand, the V-cycle is fast. 
The multigrid iterations, Algorithm~\ref{alg.MGI}, apply relaxation steps before and after the coarse grid correction. We use Gauss--Seidel iterations for this. Alternatives are schemes such as Jacobi iterations or conjugate gradient iterations. The number of Gauss--Seidel iterations applied in the pre and post relaxations is specified by the parameters $\nu_{pre}$ and $\nu_{post}$. For $V$-cycles, one sets $\nu_{pre}=\nu_{post}$. 

The norm used for the convergence test in line 9 of Algorithm~\ref{alg.MGI} depends on the context. A common choice is the standard norm of $\mathbb{R}^n$. For the Poisson and smoothing problems, we use a mass-weighted 2-norm~\cite{Wardetzky2007}. 


\section{Hierarchy Construction}
Our goal is to design a hierarchy construction that is faster than the intrinsic multigrid method by \citet{Liu2021}. It should be compatible with point clouds and general surface representations, while maintaining fast convergence during solving. Before giving an overview of our method, we revisit the idea that guided our design.

In the scheme of \citet{Liu2021}, each level is represented by a mesh and mappings to adjacent levels. An intrinsic multigrid approach can alternatively represent levels by multiple \textit{intrinsic} triangulations of the same surface. 
For example, each level can be a point sampling with corresponding intrinsic Delaunay triangulation. 
This idea, however, does not reflect a fast and more general construction. 
To achieve this, we transfer the idea into a point-cloud setting. 

\subsection{Overview}
Our approach takes as input a set of point locations $V$ sampled from a surface and a set of edges $E$ between these points denoting local neighborhoods. For a mesh $\{V, E, F\}$, we use the vertices $V$ and edges $E$. For a point cloud, edges could be taken from a radius graph or the 1-ring in a local Delaunay triangulation~\cite{Sharp2020}.

The algorithm outputs a sequence of sparse prolongation matrices $P_l$, mapping signals from $n_{l + 1}$ points to $n_l$ points, where $n_{l + 1} < n_l$. Relating $n_{l+1}$ and $n_{l}$, we refer to points in level $n_{l + 1}$ as \textit{coarse points} and points in $n_l$ as \textit{fine points}. Level $1$ are the input points.

\paragraph{Algorithm overview}
The hierarchy is constructed one level at a time. For each level, the algorithm takes the input graph from level $l$, $\{V_l, E_l\}$, and outputs the graph in level $l+1$, $\{V_{l+1}, E_{l+1}\}$. The algorithm also produces the prolongation matrix $P_l$.
This is repeated until level $\lambda$ is reached.
Here, we describe one such step. \autoref{fig:teaser} provides a corresponding visual overview.

First, the point cloud is subsampled using a fast greedy algorithm that aims to enforce a minimum edge length in the next graph (\autoref{sec:sampling}). Next, we create a graph Voronoi diagram, where the coarse points (sampled points) act as Voronoi centers and the fine points are the loci of the Voronoi cells. We then seek a mapping from the coarse points to the fine points. 
Mimicking the construction of a Delaunay triangulation as the dual of a Voronoi diagram, we construct a neighbor relation of the graph Voronoi cells (\autoref{sec:neighborgraph}) and compute all the triangles formed by the edges between Voronoi cell centers (\autoref{sec:prolongation}).
Finally, the fine points are projected onto these triangles to find the triangle closest to the fine point.
The neighbor relations of the graph Voronoi diagram are then used as edges for the next level $E_{l+1}$.

\subsection{Sampling}
\label{sec:sampling}
Each new level contains fewer points than the previous and the samplings should be spatially uniform \cite{Liu2021, Shi2006}. In other words, we seek a dense sample set $S$, in which no pair of points is closer than a prescribed distance $r$. To find such a set, we use an algorithm based on the maximum independent set: we sweep once over $V$, keeping track if points are eligible for addition to $S$. Initially, all points are eligible. If a point $p$ is eligible, we add it to $S$ and mark the points within geodesic distance $r$ of $p$ as ineligible. The radius $r$ is based on the fraction $\phi < 1$ of points we wish to keep and the average edge length $\hat{e}$
\begin{equation}
r = \phi^{-\frac{1}{3}}\hat{e}.
\end{equation}
In our experiments, we set $\phi = 1/8$ and stop coarsening at \num{1000} points, yielding roughly $\log_8(N / 1000)$ levels. We also limit the search for nearby points to the 2-ring, as this strikes a good balance between construction speed and sampling quality. In \autoref{sec:ablationsdecay} we experimentally validate that we indeed approximately reach the desired fraction of samples and \autoref{fig:levelsshowcase} demonstrates the uniformity of the resulting sampling and corresponding triangles.

\subsection{Neighbor graph}
\label{sec:neighborgraph}
We use graph Voronoi diagrams \cite{Erwig2000} to define neighborhoods for the sampled points. Since we build the levels successively from fine to coarse, the neighbor graph $\{V_l,E_l\}$ is already built on level $l$ when level $l+1$ is visited. 
The points $V_{l+1}$ are the seeds of the graph Voronoi diagram in $\{V_l,E_l\}$. For each seed $i\in V_{l+1}$, the Voronoi cell consists of the points in $V_l$ that are closer in graph distance to $i$ than to all other points of $V_{l+1}$. 
The graph Voronoi diagram can be efficiently computed by a multisource Dijkstra algorithm.
For points $i,j \in V_{l+1}$, we add an edge $\{i,j\}$ to $E_{l+1}$, if there is an edge in $E_l$ that connects a point of the Voronoi cell of $i$ with a point of the Voronoi cell of $j$. 

\subsection{Prolongation}
\label{sec:prolongation}
Prolongation operators map functions on level $l+1$ to functions on level $l$, by matrices $P_l\in \mathbb{R}^{n_l\times n_{l+1}}$.
The restrictions, mapping from level $l$ to level $l+1$, are given as the transpose matrices $P_l^{\top}$.
Important for the design of the prolongation matrices is their sparsity. The sparser the prolongation matrices, the sparser the restricted matrices $P_l^{\top}AP_l$, and the faster the mappings between the levels. To construct the prolongation, we use linear interpolation in triangles. Hereby, we get very sparse prolongations matrices. Other interpolation methods, such as radial basis functions or spline interpolations, would result in much denser prolongation matrices.

First, a set of candidate triangles on the coarse points is constructed. Every coarse point has a Voronoi cell on the finer level. Two coarse points $i, j$ are connected by an edge $\{i, j\} \in E_{l+1}$ in the coarser level if their corresponding Voronoi cells are neighbors. We consider all triangles that can be constructed from these edges:
all triplets $\{i, j, k\}$ such that $\{i, j\}, \{j, k\}, \{k, i\} \in E_{l + 1}$.

\begin{figure*}[tbh]
    \centering
    \includegraphics[width=\textwidth]{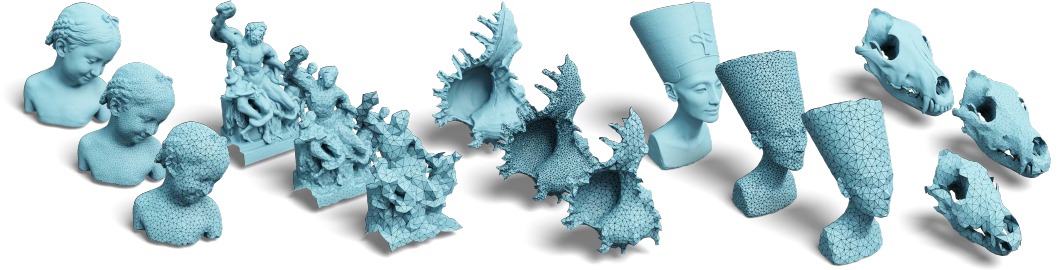}
    \caption{The input shapes and triangles in the last levels. Shapes: Bimba, Lakoon (non-manifold), Murex Romosus, Nefertiti, and Wolf Skull.}
    \label{fig:levelsshowcase}
\end{figure*}

\begin{figure*}[tbh]
    \centering
    \includegraphics[width=\textwidth]{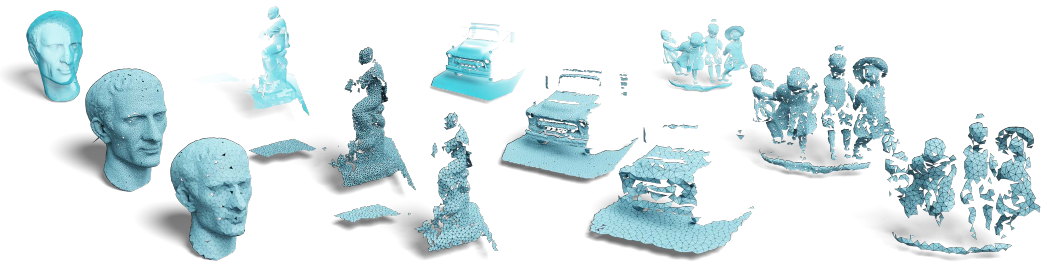}
    \caption{Input point clouds and considered triangles for the last two levels of the hierarchy. From left to right: Caesar, Ignatius, Truck, and Dancing Children.}
    \label{fig:levelsshowcase_pointcloud}
\end{figure*}

The motivation to use these edges is the duality between (intrinsic) Voronoi diagrams and Delaunay triangulations (two points in a Delaunay triangulation are connected by an edge iff their Voronoi cells are adjacent).
Since graph Voronoi cells are not continuous, but approximations computed from a sampling, the triangles we obtain are not necessarily Delaunay triangles, and they do not necessarily form a manifold. However, as illustrated in Figure~\ref{fig:levelsshowcase} and \ref{fig:levelsshowcase_pointcloud}, we mostly get well-shaped triangles and a good coverage of the surface, even for point clouds.

To get the prolongation weights for a fine point $p$, we search for the closest candidate triangle. For efficiency, we restrict this search to the triangles that include the coarse point closest to $p$. The weights are the barycentric coordinates of the closest point to $p$ in the selected triangle (this can be on an edge or a vertex).
The barycentric coordinates of the projected point are then entered into the prolongation matrix.

\paragraph{Edge-cases} In some cases, a suitable triangle cannot be found within the Voronoi neighborhood of the closest point. This might happen, for example, if all points in the neighborhood are (nearly) co-linear, or if the fine point falls outside of the triangles formed in the neighborhood. In these cases, we resort to finding the closest three points within the neighborhood and use inverse-distance weights. This is preferable over projecting to a single vertex, as it helps the spread of information during prolongation. In practice, this only happens in a fraction of cases (roughly $0.25\%$).

\paragraph{Reducing single-entry rows} The resulting prolongation matrices are very sparse with maximally three non-zero entries per row. 
Since the coarse points are created by subsampling the fine points, the fine points that are sampled transfer their function value directly to the corresponding coarse point during prolongation. Therefore, there is only one entry in the corresponding rows. 
We obtain prolongation matrices with fewer single-entry rows by moving each sampled point to the mean of the points that form its graph Voronoi cell before we compute the closest point projections. In our experiments, we obtained a slight improvement of solving times with this strategy over not moving the coarse points.

\section{Experiments}
We evaluate \ourmethod~ and compare to state-of-the-art GMG methods and AMG approaches. For reference, we provide results for direct solvers. We also provide insights into design choices via ablation studies.

\label{sect.experiments}
\subsection{Implementation}
Our multigrid-solver implementation builds on \citet{Liu2021}'s code, where the prolongation matrix definition is exchanged. In every experiment, we set the number of pre/post-relaxation steps $v_{pre} = v_{post} = 2$.
The hierarchy construction uses custom routines built around Eigen~\cite{eigenweb} and only requires a matrix of points and an array of edges. 
The code for our solver is available as a C++ library and Python package, along with scripts to replicate the main tables and figures in this paper: \url{https://graphics.tudelft.nl/gravo_mg}.

We use an Intel\textregistered Core\texttrademark i9-9900 CPU (3.10GHz, 32GB memory). The code does not employ multithreading but could be parallelized. None of the methods in our comparisons are parallelized, except \textsc{Pardiso}. A discussion on the potential for parallelization of the solver we used can be found in Section 7 of \cite{Liu2021}.

\subsection{Problems}
In our ablations and comparisons, we test our approach on two standard problems that can be written as linear systems: data smoothing and Poisson problems.
For meshes, both problems involve the cotan Laplace matrix $S$ and the lumped mass matrix $M$, see \cite{Wardetzky2007}. In the case of point clouds and non-manifold meshes, we use the robust Laplacian by \citet{Sharp2020}.
Data can be smoothed by solving
\begin{equation}\label{eq.smoothing}
(M + \alpha S)x = My,
\end{equation}
where $y$ is the noisy input function and $\alpha$ a parameter that determines how much the data is smoothed. 
The Poisson problem is 
\begin{equation}\label{eq.Poisson}
(S + \eta M)x = My,
\end{equation}
where $y$ is a random vector. The term $\eta M$ is added to obtain a positive-definite system matrix. The parameter $\eta$ is chosen to be very small, for example $\eta=\num{1e-6}$.
Our solver terminates when tolerance $\varepsilon$ is reached (line 9 of \autoref{alg.MGS}) or after a maximum number of iterations.

\begin{table}[b]
\caption{Data smoothing timings with variations of the sampling step (\textit{Smp}). We compare our approach to \textit{random} sampling, Poisson Disk Sampling (\textit{PDS}), geodesic Farthest Point Sampling (\textit{FPS}), and Maximum-Independent Set (\textit{MIS}). All timings are in seconds, unless otherwise specified.}
\centering
\resizebox{\columnwidth}{!}{
\begin{tabular}{@{}lr|rr||rr|rr|rr|rr@{}}
\hline

 \multicolumn{1}{c}{\multirow{2}{*}{\textbf{Model}}} & \multicolumn{1}{c|}{\multirow{2}{*}{\textbf{\#V}}} & \multicolumn{2}{c||}{\textbf{Ours}} & \multicolumn{2}{c|}{\textbf{Random}} & \multicolumn{2}{c|}{\textbf{PDS}} & \multicolumn{2}{c|}{\textbf{FPS}} & \multicolumn{2}{c}{\textbf{MIS}} \\
\cline{3-12}
\multicolumn{1}{c}{} & \multicolumn{1}{c|}{} & \small{Smp} & \small{Solve}  & \small{Smp} & \small{Solve} & \small{Smp} & \small{Solve} & \small{Smp} & \small{Solve} & \small{Smp} & \small{Solve} \\
\hline
 Brd Man      & 691k  &        0.08 &        \textbf{0.62} &         0.01 &        1.01 &         0.47 &        0.63  &        1m &        0.79 &         0.02 &        0.87 \\
 Rd Circ. Box & 701k  &        0.08 &        \textbf{0.81} &         0.01 &        1.30 &         0.39 &        1.10  &        1m  &       1.38 &         0.02 &        1.01 \\
 Nefertiti    & 1m    &        0.12 &        \textbf{1.10} &         0.01 &        2.05 &         1.02 &        1.29  &        2m &        1.65 &         0.04 &        1.15 \\
 Murex        & 1.8m  &        0.42 &        \textbf{3.02} &         0.03 &        4.84 &         1.47 &        3.41  &        6m &        4.83 &         0.11 &        3.68 \\
 XYZ Dragon   & 3.6m  &        0.35 &        5.72 &         0.06 &       12.95 &         2.48 &        \textbf{4.92}  &       27m &           - &         0.11 &        7.38 \\
\hline
\end{tabular}
}
\label{tab:ablations_sampling}
\end{table}
\begin{table*}[tb]
\caption{Timings for hierarchy construction and solving on data smoothing with variations of the entries in the prolongation operator. \textit{Ours} only considers triangles formed by Voronoi edges. The other variants either pick $n$ \textit{closest} points, pick $n$ \textit{random} points or pick the three Voronoi neighbors that form the \textit{closest triangle}. All timings are in seconds.}
\centering
\resizebox{0.9\textwidth}{!}{
\begin{tabular}{@{}lr|rrr||rrr||rrr|rrr|rrr||rrr@{}}
\hline
 \multicolumn{1}{c}{\multirow{3}{*}{\textbf{Model}}} & \multicolumn{1}{c|}{\multirow{3}{*}{\textbf{\#Vert}}} & \multicolumn{3}{c||}{\textbf{3 points}} & \multicolumn{3}{c||}{\textbf{2 points}} & \multicolumn{9}{c||}{\textbf{3 points}} &  \multicolumn{3}{c}{\textbf{4 points}} \\
 \cline{3-20}
 \multicolumn{1}{c}{} & \multicolumn{1}{c|}{} & \multicolumn{3}{c||}{\textbf{Ours}} & \multicolumn{3}{c||}{\textbf{Closest}} & \multicolumn{3}{c|}{\textbf{Random}} & \multicolumn{3}{c|}{\textbf{Closest vert}} & \multicolumn{3}{c||}{\textbf{Closest tri}} & \multicolumn{3}{c}{\textbf{Closest}} \\
\cline{3-20}
\multicolumn{1}{c}{} & \multicolumn{1}{c|}{} & \small{Hier} & \small{\#It} & \small{Solve} & \small{Hier} & \small{\#It} & \small{Solve} & \small{Hier} & \small{\#It} & \small{Solve} & \small{Hier} & \small{\#It} & \small{Solve} & \small{Hier} & \small{\#It} & \small{Solve} & \small{Hier} & \small{\#It} & \small{Solve}   \\
\hline
 Beard Man        & 691k  &        0.64 &        4 &        \textbf{0.62} &        0.46 &        9 &        0.91 &        0.42 &       26 &        2.91 &        0.50 &        5 &        0.72 &        0.87 &        5 &        0.77 &        0.51 &        5 &        0.83 \\
 Red Circular Box & 701k  &        0.67 &        6 &        \textbf{0.79} &        0.49 &       10 &        1.03 &        0.49 &       31 &        3.45 &        0.52 &        7 &        0.95 &        1.01 &        8 &        1.06 &        0.58 &        7 &        1.01 \\
 Nefertiti        & 1m    &        0.93 &        4 &        \textbf{1.05} &        0.67 &        9 &        1.68 &        0.60 &       29 &        5.75 &        0.69 &        6 &        1.36 &        1.35 &        5 &        1.29 &        0.73 &        5 &        1.37 \\
 Murex Romosus    & 1.8m  &        2.64 &        5 &        \textbf{3.11} &        1.80 &       11 &        4.15 &        1.70 &       40 &       15.07 &        2.07 &        6 &        3.60 &        3.82 &        6 &        3.41 &        1.97 &        6 &        3.91 \\
 XYZ Dragon       & 3.6m  &        3.46 &        9 &        \textbf{5.72} &        2.41 &       15 &        7.55 &        2.28 &       45 &       25.10 &        2.61 &       14 &        8.35 &        4.45 &       16 &        8.76 &        2.68 &       11 &        7.32 \\
\hline
\end{tabular}
}
\label{tab:ablations_triangleselection}
\end{table*}

\subsection{Ablation Studies}
We would like to understand the effects of our design choices on the hierarchy construction and subsequent solving steps. We structure these experiments along three themes: sampling, prolongation selection, and weighting. In each ablation, we compare variants of our approach on a fixed set of meshes and point clouds and run a data smoothing problem as detailed above. We smooth a random function with $\alpha=$ \num{1e-3} and tolerance \num{1e-4}. Each variant is then evaluated in terms of time to construct the hierarchy, the number of iterations required to reach  the target tolerance, and the total time.

\begin{figure}[b]
    \centering
    \includegraphics[width=0.7\columnwidth]{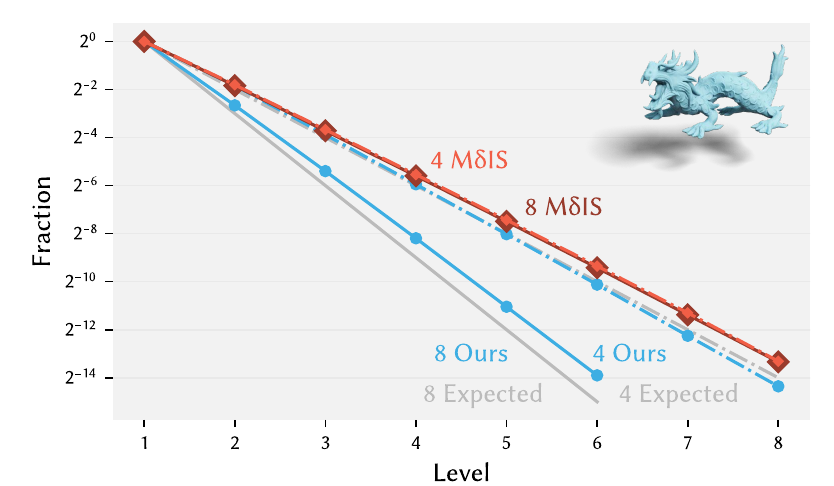}
    \caption{Comparison of the decay rate between M$\delta$IS used by \citet{Shi2006} and our approach on XYZ dragon. The y-axis is in $\log_2$ scale.}
    \label{fig:decayplot}
\end{figure}

\paragraph{Sampling}
\label{sec:ablationsdecay}
We seek a sampling method that balances run-time during hierarchy construction and sampling quality. To validate that our approach effectively balances these demands, we compared our approach to random sampling, Poisson-disk-sampling (PDS), geodesic farthest point sampling (FPS), and maximal independent set (MIS) selection. For every method, we set the target ratio between levels to 1/8th.
In \autoref{tab:ablations_sampling}, we observe that our approach is faster than the others when solving. When considering the full hierarchy construction, our approach is faster than every other sampling approach, because we perform part of the graph Voronoi diagram construction during sampling and have fewer points per level to consider than MIS.

We also compare decay rate between the maximal $\delta$-independent set, used in \cite{Shi2006}, and our sampling. In \autoref{fig:decayplot}, we see that it is possible to decay much faster with our approach than M$\delta$IS, because it must always be a superset of the MIS. This is an advantage of our approach; we can perform faster and fewer iterations, while having a fast sampling time.

\begin{figure}[b]
    \centering
    \includegraphics[width=0.7\columnwidth]{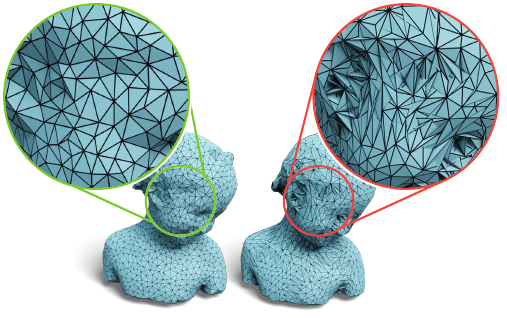}
    \caption{Using dual Voronoi triangles results in a more consistent set of candidate triangles than using all triangles in the coarse point's 1-ring.}
    \label{fig:nonvoronoi}
\end{figure}

\paragraph{Prolongation selection} Our approach uses triangles of coarse points for the prolongation operator. This results in sparse prolongation matrices that spread information in each tangential direction. In this ablation, we seek to support this choice. In \autoref{tab:ablations_triangleselection}, we compare our approach to the following variants that do not explicitly work with triangles: 
simply prolonging to the closest two, three, or four points from the graph Voronoi neighbors and picking three random points. We also test a variant that considers triangles without restricting to Voronoi edges, `closest tri'. This results in less consistent triangulations, as shown in \autoref{fig:nonvoronoi}. For each of the non-triangle selection approaches, we use inverse distance weights. We observe that our approach solves faster than all other variants, while increasing the hierarchy construction time only a bit.

\begin{table}[b]
\caption{Timings for solving on data smoothing with variations of the weighting scheme. We compare barycentric coordinates (\textit{Ours}) to \textit{uniform} weights, \textit{inv}erse \textit{dist}ance weights, and barycentric coordinates without chaging the positions of the coarse points before projection (\textit{No shift}). All timings are in seconds.}
\centering
\resizebox{\columnwidth}{!}{
\begin{tabular}{@{}lr|rr||rr|rr|rr@{}}
\hline

 \multicolumn{1}{c}{\multirow{2}{*}{\textbf{Model}}} & \multicolumn{1}{c|}{\multirow{2}{*}{\textbf{\#Vert}}} & \multicolumn{2}{c||}{\textbf{Ours}} & \multicolumn{2}{c|}{\textbf{Uniform}} & \multicolumn{2}{c|}{\textbf{Inv. Dist.}} & \multicolumn{2}{c}{\textbf{No shift}}  \\
\cline{3-10}
\multicolumn{1}{c}{} & \multicolumn{1}{c|}{} & \small{\#It} & \small{Solve} & \small{\#It} & \small{Solve} & \small{\#It} & \small{Solve} & \small{\#It} & \small{Solve}  \\
\hline
 Beard Man        & 691k  &        4 &        0.63 & 12 &        1.30 &  5 &        0.71 &        4 &        \textbf{0.56} \\
 Red Circular Box & 701k  &        6 &        \textbf{0.80} & 23 &        2.36 &  6 &        \textbf{0.80} &       10 &        1.12 \\
 Nefertiti        & 1m &        4 &        1.04 & 14 &        2.68 &  5 &        1.21 &        4 &        \textbf{0.97} \\
 Murex Romosus    & 1.8m &        5 &        3.09 & 16 &        6.54 &  6 &        3.37 &        5 &        \textbf{2.71} \\
 XYZ Dragon       & 3.6m &        9 &        \textbf{5.71} & 23 &       12.23 &  9 &        5.74 &       18 &        9.28 \\
\hline
\end{tabular}
}

\label{tab:ablations_weighting}
\end{table}

\paragraph{Weighting} We project the fine points onto the triangles formed by coarse points and use barycentric weights as predictors for the value of the fine point. Previous works suggest that the choice of weighting schemes has little effect on convergence times \cite{Aksoylu2005, Shi2006}, while \citet{Liu2021} argue that the weighting scheme is crucial for some shapes. In \autoref{tab:ablations_weighting}, we compare our approach with uniform weights and inverse distance weights alongside a variant where we do not shift the coarse points to the barycenters. We observe that our approach works best with barycentric coordinates (\textit{Ours}). Inverse-distance weights are not far behind. Shifting coarse points has benefits for some, but not all shapes. This is not the core contribution of our work and could be left out in some cases. A benefit of not shifting coarse points is that each iteration is faster because the prolongation matrix contains more single-entry rows.

\begin{table*}[h]
\caption{Comparison of our hierarchy construction and solver for a Poisson problem with $\eta=$\num{1e-6} mass matrix coefficient and tolerance of \num{1e-4}. Missing entries are not available for the given method. The maximum number of iterations for iterative solvers is set to 100.}
\centering
\resizebox{\textwidth}{!}{
\begin{tabular}{@{}lr|rrr|rrr|rrr||rrr|rrr||rr|rr@{}}
\hline

 \multicolumn{1}{c}{\multirow{2}{*}{\textbf{Model}}} & \multicolumn{1}{c|}{\multirow{2}{*}{\textbf{\#Vert}}} & \multicolumn{3}{c|}{\textbf{\ourmethod~(Ours)}} & \multicolumn{3}{c|}{\textbf{\citet{Liu2021}}} & \multicolumn{3}{c||}{\textbf{\citet{Shi2006}}} & \multicolumn{3}{c|}{\textbf{AMG-RS}} & \multicolumn{3}{c||}{\textbf{AMG-SA}} & \multicolumn{2}{c|}{\textbf{Eigen}} & \multicolumn{2}{c}{\textbf{\textsc{Pardiso}}}  \\
\cline{3-21}
\multicolumn{1}{c}{} & \multicolumn{1}{c|}{} & \small{Hier} & \small{\#It} & \small{Solve} & \small{Hier} & \small{\#It} & \small{Solve} & \small{Hier} & \small{\#It} & \small{Solve} & \small{Hier} & \small{\#It} & \small{Solve} & \small{Hier} & \small{\#It} & \small{Solve} & \small{Fact.} & \small{Subst.} & \small{Fact.} & \small{Subst.}   \\
\hline
\multicolumn{19}{c}{\small{\textsc{manifold triangular meshes}}} \\
\hline
 Aim Dragon       & 152k  &        0.14 &        7 &        0.19 &              5.14 &        8 &        0.27 &              0.06 &       27 &        0.62 &           0.15 &       26 &        0.46 &           0.31 &       29 &        0.48 &             0.81 &              0.02 &             0.58 &              0.04 \\
 Blade Smooth     & 195k  &        0.15 &        4 &        0.17 &              5.87 &        3 &        0.17 &              0.09 &       18 &        0.60 &           0.18 &      100 &        2.42 &           0.40 &       42 &        0.94 &             0.76 &              0.02 &             0.69 &              0.04 \\
 Moses            & 258k  &        0.42 &       12 &        0.55 &              8.72 &        5 &        0.35 &              0.30 &      100 &        4.30 &           0.27 &      100 &        3.64 &           0.55 &      100 &        3.33 &             0.90 &              0.02 &             1.04 &              0.05 \\
 Julius Caesar    & 387k  &        0.30 &       11 &        0.58 &             12.10 &       17 &        1.09 &              0.16 &       28 &        1.54 &           0.39 &      100 &        4.89 &           0.77 &       70 &        2.79 &             5.07 &              0.06 &             1.29 &              0.09 \\
 Bimba            & 502k  &        0.43 &        7 &        0.65 &             15.58 &        6 &        0.74 &              0.24 &       48 &        4.16 &           0.51 &      100 &        6.97 &           1.15 &       69 &        4.45 &             3.54 &              0.05 &             2.13 &              0.10 \\
 Antique Head     & 651k  &        0.53 &        4 &        0.51 &             20.07 &        4 &        0.60 &              0.28 &       14 &        1.22 &           0.64 &      100 &        8.48 &           1.37 &       67 &        4.33 &            15.02 &              0.11 &             2.43 &              0.17 \\
 Beard Man        & 691k  &        0.59 &        4 &        0.56 &             22.15 &        3 &        0.58 &              0.26 &       14 &        1.43 &           0.52 &      100 &        7.07 &           1.46 &       14 &        0.96 &            24.57 &              0.14 &             2.72 &              0.19 \\
 Red Circular Box & 701k  &        0.64 &        6 &        0.74 &             22.97 &        6 &        0.97 &              0.34 &       66 &        6.67 &           0.72 &      100 &        8.98 &           1.51 &       66 &        5.01 &            17.76 &              0.11 &             2.84 &              0.17 \\
 Dancing Children & 724k  &        0.69 &        9 &        1.10 &             23.21 &        8 &        1.25 &              0.41 &       39 &        4.54 &           0.68 &      100 &        9.32 &           1.23 &      100 &        8.70 &             6.18 &              0.09 &             2.78 &              0.17 \\
 Ramses           & 826k  &        0.79 &        7 &        1.21 &             28.90 &        5 &        1.18 &              0.47 &       40 &        6.03 &           0.91 &      100 &       11.82 &           2.08 &       49 &        5.40 &             6.24 &              0.09 &             3.60 &              0.18 \\
 Nefertiti        & 1m &        0.89 &        4 &        0.94 &             34.22 &        4 &        1.18 &              0.56 &       65 &       11.69 &           1.09 &      100 &       14.74 &           2.58 &       46 &        6.14 &             9.15 &              0.10 &             4.54 &              0.22 \\
 Isidore Horse    & 1.1m &        1.12 &       11 &        1.90 &             35.60 &        5 &        1.27 &              0.50 &       70 &       11.03 &           1.11 &      100 &       14.60 &           2.50 &       88 &       10.72 &            24.01 &              0.17 &             4.56 &              0.28 \\
 Ram              & 1.3m &        2.40 &        3 &        1.95 &             56.18 &        - &        - &              1.14 &       49 &       13.39 &           2.45 &      100 &       25.95 &           4.72 &      100 &       21.71 &            17.07 &              0.15 &             6.99 &              0.33 \\
 Murex Romosus    & 1.8m &        2.32 &        6 &        2.85 &             73.05 &        5 &        3.32 &              0.99 &       63 &       20.79 &           2.38 &      100 &       29.83 &           5.08 &       58 &       15.28 &            40.06 &              0.26 &             9.13 &              0.44 \\
 XYZ Dragon       & 3.6m &        3.24 &        9 &        5.32 &            121.97 &        7 &        5.32 &              1.57 &       55 &       28.95 &           3.16 &      100 &       43.75 &           9.14 &       75 &       30.54 &            77.62 &              0.69 &            15.88 &              0.94 \\
\hline
\multicolumn{19}{c}{\small{\textsc{non-manifold triangular meshes}}} \\
\hline
 Lakoon            & 188k  &        0.16 &        8 &        0.29 & - & - & - &              0.09 &       41 &        1.33 &           0.21 &      100 &        2.77 &           0.47 &       48 &        1.12 &             0.42 &              0.01 &             0.71 &              0.04 \\
 Indonesian Statue & 294k  &        0.26 &       11 &        0.58 & - & - & - &              0.16 &       64 &        3.15 &           0.30 &      100 &        3.92 &           0.63 &      100 &        3.55 &             0.92 &              0.03 &             1.18 &              0.06 \\
 Beethoven         & 383k  &        0.45 &        4 &        0.49 & - & - & - &              0.23 &       60 &        4.00 &           0.51 &       20 &        1.29 &           0.96 &      100 &        5.30 &             2.39 &              0.04 &             1.65 &              0.09 \\
 Bayon Lion        & 749k  &        1.42 &        6 &        1.55 & - & - & - &              0.70 &       26 &        4.31 &           1.30 &      100 &       15.36 &           2.49 &       43 &        5.57 &             5.99 &              0.08 &             3.79 &              0.18 \\
 Helmet Moustache  & 941k  &        2.04 &        9 &        2.89 & - & - & - &              0.74 &       57 &       11.15 &           2.03 &      100 &       19.66 &           3.31 &       38 &        6.14 &            24.66 &              0.14 &             5.56 &              0.25 \\
 Zeus              & 1.3m &        2.47 &       11 &        3.86 & - & - & - &              1.17 &       58 &       15.91 &           2.34 &      100 &       27.21 &           4.11 &      100 &       22.68 &            30.40 &              0.20 &             7.19 &              0.35 \\
 Alfred Jacquemart & 1.4m &        3.33 &        5 &        3.79 & - & - & - &              1.67 &       43 &       16.21 &           3.03 &      100 &       30.33 &           5.44 &       51 &       12.78 &             8.88 &              0.14 &             8.07 &              0.35 \\
\hline
\multicolumn{19}{c}{\small{\textsc{point clouds}}} \\
\hline
 Oil Pump      & 103k  &        0.07 &        9 &        0.12 & - & - & - &              0.04 &       15 &        0.22 &           0.10 &      100 &        1.27 &           0.19 &       55 &        0.56 &             0.17 &              0.01 &             0.31 &              0.02 \\
 Caesar Merged & 388k  &        0.29 &        6 &        0.40 & - & - & - &              0.17 &       18 &        1.14 &           0.41 &      100 &        5.52 &           0.83 &       87 &        3.98 &             4.90 &              0.06 &             1.50 &              0.10 \\
 Truck         & 1.2m &        0.99 &       17 &        3.20 & - & - & - &              0.65 &       26 &        5.53 &           1.27 &      100 &       18.87 &           3.67 &       72 &       10.77 &             5.63 &              0.14 &             5.09 &              0.29 \\
 Ignatius      & 1.4m &        1.26 &        7 &        1.87 & - & - & - &              0.78 &       33 &        8.41 &           1.59 &      100 &       21.61 &           4.42 &      100 &       17.78 &             8.92 &              0.18 &             6.13 &              0.36 \\
\hline
\end{tabular}
}

\label{tab:comparisonpoisson}
\end{table*}

\subsection{Comparisons}
We compare our approach on a wide range of meshes and point clouds for a Poisson problem with $\eta=$\num{1e-6} and target tolerance of \num{1e-4}.
The input function $y$ is a random vector sampled from $\mathcal{N}(0, 1)$.

\begin{figure*}[tbh]
    \centering
    \includegraphics[width=\textwidth]{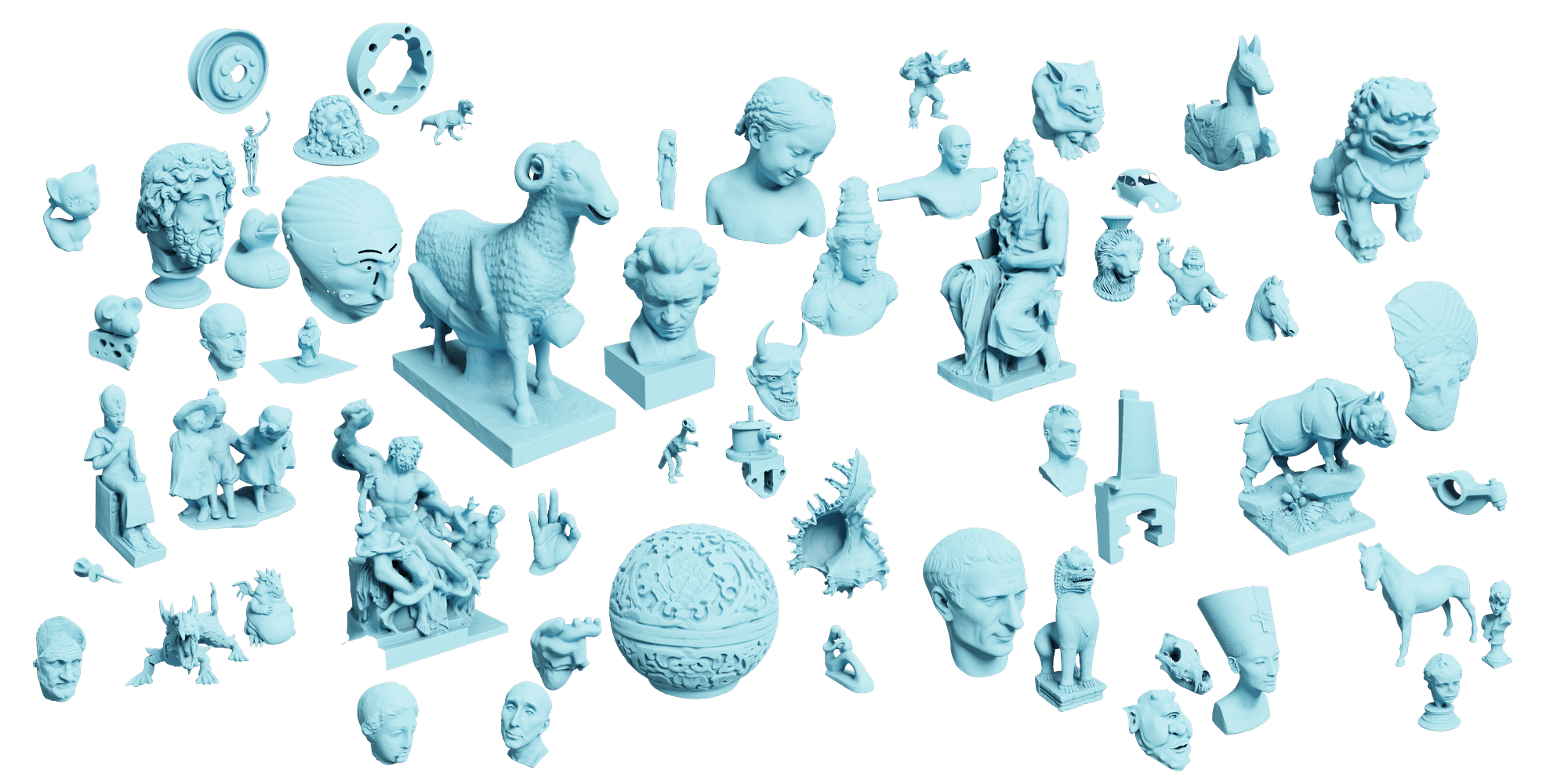}
    \caption{All (non)manifold triangular meshes used in our experiments. Mosaic generated with code from Qingnan Zhou.}
    \label{fig:allmeshes}
\end{figure*}

The shapes were selected to have at least 100k vertices and exhibit a wide variety: uniform meshes (\textit{e.g.}, Nefertiti), non-uniform meshes (\textit{e.g.}, Alfred Jacquemart, Indonesian statue), broken and non-manifold meshes. The meshes also exhibit detailed features (\textit{e.g.}, XYZ dragon) and complex curvature (\textit{e.g.}, Murex Romosus). All the shapes
are shown in \autoref{fig:allmeshes}. We make no use of additional pre-processing steps, such as remeshing or fixing non-manifold edges: every mesh is used as-is in the highest resolution available from the respective sources. For the point clouds, we opted for high-resolution scanned data. The point clouds come from the Tanks and Temples benchmark dataset~\cite{Knapitsch2017}
and from range scans in the AIM@Shape repository~\cite{aimatshape}.

\ourmethod~is compared to the GMG solvers by \citet{Liu2021} and \citet{Shi2006}, and the AMG methods Ruge--Stuben and Smoothed Aggregation. For reference, we list the timings of direct solvers. For \citeauthor{Liu2021}, we use their provided implementation. We reimplemented  \citeauthor{Shi2006} based on their paper. The latter mentions multiple weighting schemes, including uniform weights and inverse distance weights. We tested both and report the best-performing approach: inverse distance weights. For the AMG approaches, we use the implementation provided in PyAMG~\cite{BeOlSc2022} with default settings provided by the package. We set the maximum number of iterations for all iterative solvers to 100, since more iterations would not change the overall picture regarding which solver is faster. The direct solver references are the Cholesky LLT factorization provided in Eigen and Intel\textregistered MKL's \textsc{Pardiso} solver, which is highly optimized and parallelized~\cite{intel_onemkl_pardiso}.

Our approach yields faster solving times for the majority of input meshes (\autoref{tab:comparisonpoisson}). More results for manifold meshes are listed in the supplement in Table 1. On average, our construction is 36x faster than \citeauthor{Liu2021} and only 1.8x slower than \citeauthor{Shi2006}'s method. With regards to solving time, \citeauthor{Liu2021} takes $3\%$ more time on average for the Poisson problem and $7\%$ for data smoothing and \citeauthor{Shi2006} takes $274\%$ more time for the Poisson problem and $81\%$ for data smoothing. Note that we require less time for one iteration than  \citeauthor{Liu2021}, because we use a higher decay rate (1/8 vs. 1/4). This is balanced out in most cases by a higher iteration count and the overall solving times are similar when we use a decay rate of 1/4.

GMG methods are most beneficial in settings where a user would iterate on the system matrix, but the benefit of using \ourmethod~is already noticeable starting with the first solve. For all meshes larger than 100k vertices, our approach is faster than \citeauthor{Liu2021} for both the Poisson problem and data smoothing. The same holds for \citeauthor{Shi2006} for the Poisson problem. For data smoothing, we are faster for one solve in $83\%$ of cases and for three solves in $93\%$ of cases. Compared to the \textsc{Pardiso} solver, we are faster for one solve of the Poisson problem in $92\%$ of cases and for three solves in $95\%$ of cases (data smoothing, 1x: $95\%$, 3x: $98\%$). Note, however, that our solver stops at a higher residual error than direct solvers. The strength of multigrid approaches is in settings where one needs a quick and relatively accurate solution. A direct solver is often preferable in settings where high accuracy is required.

\begin{figure}[htb]
    \centering
    \includegraphics[width=0.8\columnwidth]{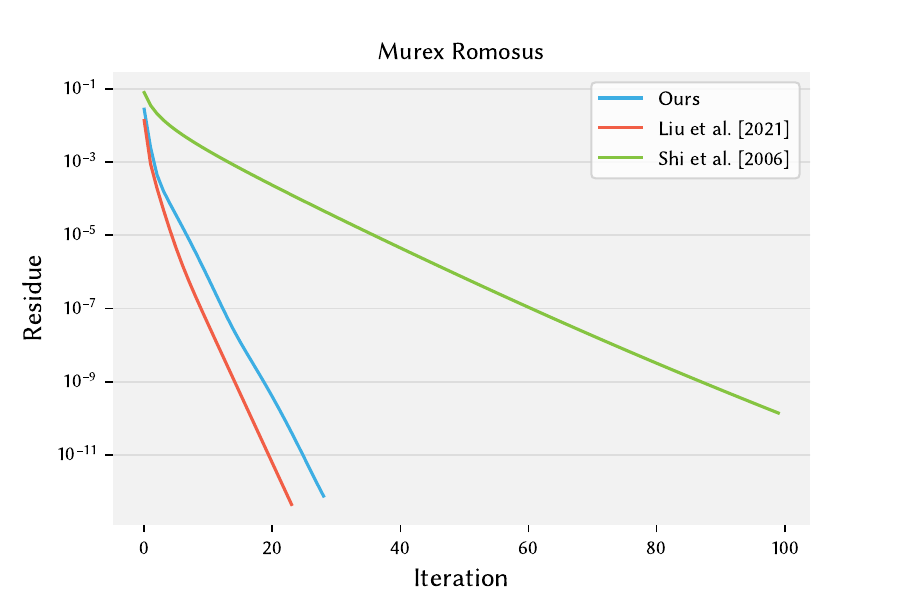}
    \caption{Residual over iteration count for Ours, \citet{Liu2021}, and \citet{Shi2006} for data smoothing on Murex Romosus with $\alpha = $\num{1e-3}. See \autoref{fig:teaser} for a plot over time.}
    \label{fig:convergenceplot}
\end{figure}

To provide insight into the convergence of our approach compared to the other GMG schemes, we plot convergence for a data smoothing problem with $\alpha=$\num{1e-3} for the Murex Romosus shape in \autoref{fig:teaser} and the same plot against number of iterations in \autoref{fig:convergenceplot}. Again we see that our approach is on par with \citet{Liu2021} and beats \citet{Shi2006} with a high margin. More convergence plots for data smoothing, including plots over the number of iterations, can be found in the supplement. These confirm our results. There are some outliers: for Red Circular Box, \citet{Shi2006} converges faster than the other GMG approaches and for Moses, \ourmethod~slows down around a residual of \num{1e-6}.

\subsection{Applications}
\label{sec:applications}
We evaluate our solver in three scenarios: data smoothing, a geometric flow, and physical simulation. We compare solving times to a sparse Cholesky solver, commonly used for these problems. 

\begin{figure}[tbh]
  \centering
  \includegraphics{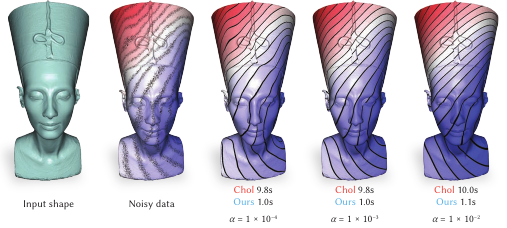}
  \parbox[t]{1.0\columnwidth}{\relax  }
  \caption{\label{fig:smooting-mesh}Smoothing of scalar data on a surface mesh with various parameter settings (Model: Nefertiti, 1m vertices) using the Dirichlet energy as smoothness energy.
}
\end{figure}

\begin{figure}[tbh]
  \centering
  \includegraphics{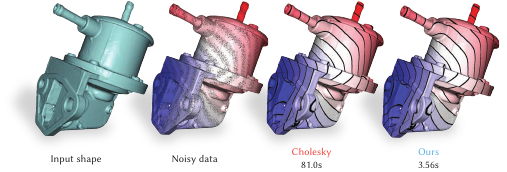}
  \parbox[t]{1.0\columnwidth}{\relax  }
  \caption{\label{fig:smoothing-bilaplacian}Smoothing of scalar data on a surface mesh (Model:Oilpump, 570k vertices) using a weighted sum of the Dirichlet energy and a bi-Laplacian energy as smoothness energy.
}
\end{figure}

\paragraph{Data smoothing}
For data smoothing, we consider an input function y on a surface and compute a smoother function x by minimizing a quadratic objective
\begin{equation}
(x-y)^{\top}M(x-y)+\alpha x^{\top}Sx+\beta x^{\top}SM^{-1}Sx.
\end{equation}
The first term is a data term that penalizes deviation from the input function, the second and third terms are Laplace and bi-Laplace smoothing energies and $\alpha,\beta\in\mathbb{R}^{\ge 0}$ are parameters.
Results are shown in Figures \ref{fig:smooting-mesh} and \ref{fig:smoothing-bilaplacian}.
The figures list timings for solving the linear systems with our method and Eigen's sparse Cholesky solver. When changing parameter $\alpha$ to adjust the amount of smoothing, the direct solver needs to compute a new matrix factorization resulting in significant solving-time differences compared to our solver, in particular, when the bi-Laplacian energy is included.

\begin{figure}[tbh]
  \centering
\includegraphics{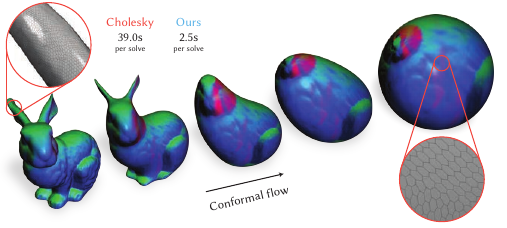}
  \parbox[t]{1.0\columnwidth}{\relax  }
  \caption{\label{fig:flowPoly} Conformal flow on polygon mesh. (Bunny model, 626k vertices).
}
\end{figure}

\begin{figure}[tbh]
  \centering
  \includegraphics{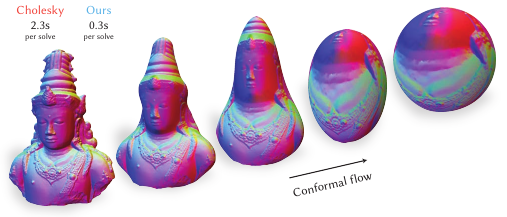}
  \parbox[t]{1.0\columnwidth}{\relax  }
  \caption{\label{fig:flowNonMan} We compare the performance of our multigrid solver against a direct solver on conformal mean curvature flow on a nonmanifold mesh. (Indonesian statue model, 295k vertices).
}
\end{figure}

\paragraph{Conformal flow} 
As an example of a nonlinear geometric flow, we consider the conformal flow~\cite{kazhdan2012can}. For robustness, we use an implicit time-integration that requires solving a linear Laplace system for every time step. 
We show results in Figures \ref{fig:flowPoly} and \ref{fig:flowNonMan} and compare our solving times to those of Eigen's sparse Cholesky solver. 
Since the system matrix changes every time step, the direct solver constantly needs to compute new factorizations, resulting in substantial differences when performing multiple steps.

\begin{figure}[tbh]
  \centering
  \includegraphics{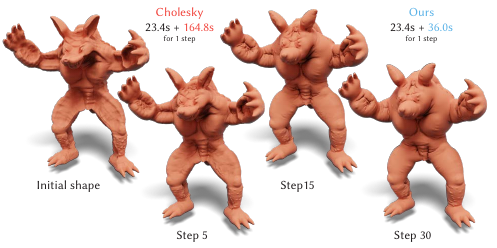}
  \parbox[t]{1.0\columnwidth}{\relax  }
  \caption{\label{fig:inflation} Simulation of balloon inflation (Model: Armadillo, 180k vertices).
}
\end{figure}

\paragraph{Balloon inflation} 
As an example of a physical simulation, we consider the balloon inflation from \cite{skouras2012computational}. A surface mesh represents a thin-layered rubber balloon that undergoes membrane deformation subject to air pressure. For time-integration an implicit Euler scheme is used and the resulting nonlinear equations are solved using a Newton scheme. To find the descent direction a sparse linear system is solved. As for the geometric flow, due the simulation's nonlinearity, the system matrix changes with every time step, forcing the direct solver to compute a new factorization in every time step. 
Results and timings are shown in Fig~\ref{fig:inflation}. 

\section{Conclusion}
We introduce \ourmethod, a surface multigrid method that features fast hierarchy construction, applicability to general surface representations, and fast convergence. 
Our experiments demonstrate excellent performance compared to other GMG and AMG methods and direct solvers. 

Conceptually, our method deviates from the common paradigm of GMG to represent levels via watertight meshes obtained by edge collapse. 
We use the geometry of the surface, while AMG ignores it for hierarchy construction. 
This opens a new direction for GMG on manifolds, which are generally applicable and fast to build, hereby improving the scalability of geometry processing methods. 

In future work, graph Voronoi diagrams could be used for point cloud processing. We are excited about the quality of the triangles we generate from the graph Voronoi diagrams and see a potential use, when fast triangulations or uniform samples on point clouds are needed. Regarding theory, it would be interesting to explore under which conditions this approach can provide guarantees regarding the triangulation. 
Further acceleration is still possible. An important aspect is parallelization of both the hierarchy construction and the solver. 
Our solver could also become a preconditioner for a Krylov method like GMRES or CG to accelerate convergence. 

\begin{acks}
We thank the anonymous reviewers for their constructive feedback. This project is partly supported by the Indonesia Endowment Fund for Education (LPDP) via a doctoral scholarship for Ahmad Nasikun and by the Directorate of Research (Direktorat Penelitian) of the Universitas Gadjah Mada (UGM).
We thank the authors of the meshes used in our experiments. A full list of meshes and attributions is listed at \url{https://graphics.tudelft.nl/gravo_mg}.
\end{acks}

\bibliographystyle{ACM-Reference-Format}
\bibliography{geometric-multigrid}

\clearpage

\clearpage
\appendix
\section{Extended comparisons}
In \autoref{tab:comparisonallpoisson} (second page), we show more results for the Poisson problem on manifold meshes. Note that some entries are listed as NaN. This arises from the Gauss--Seidel smoothing step, where a divison by the diagonal of the system matrix is performed. In the cases where a NaN arises, the restriction of the system matrix results in zero-entries on the diagonal. This is not a fundamental issue for the Gauss--Seidel solver. The issue could, for example, be addressed by including a pivoting strategy. We did not include these entries for the conclusions listed in the main paper.

\begin{figure*}
    \begin{subfigure}{0.24\textwidth}
        \centering
        \includegraphics[width=\textwidth]{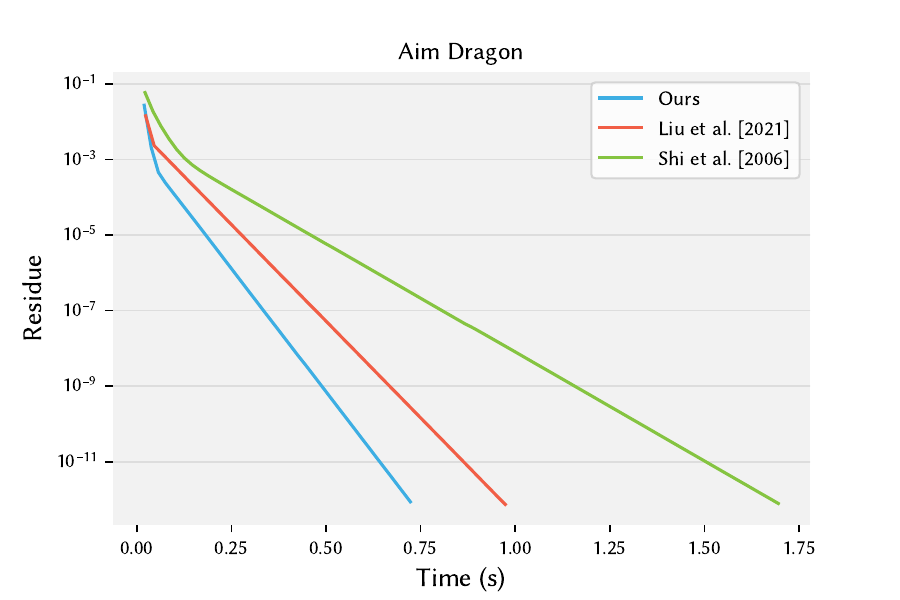}
    \end{subfigure}
    \begin{subfigure}{0.24\textwidth}
        \centering
        \includegraphics[width=\textwidth]{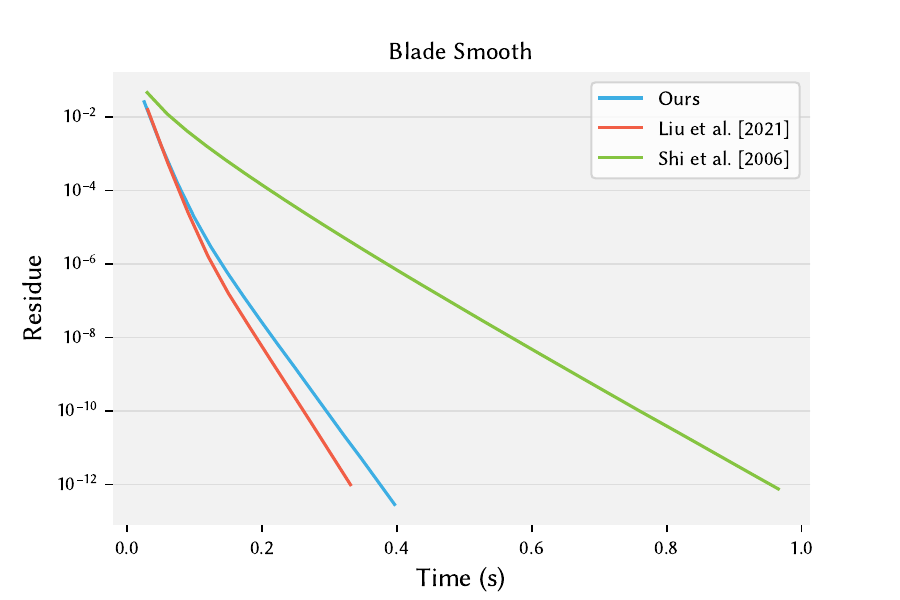}
    \end{subfigure}
    \begin{subfigure}{0.24\textwidth}
        \centering
        \includegraphics[width=\textwidth]{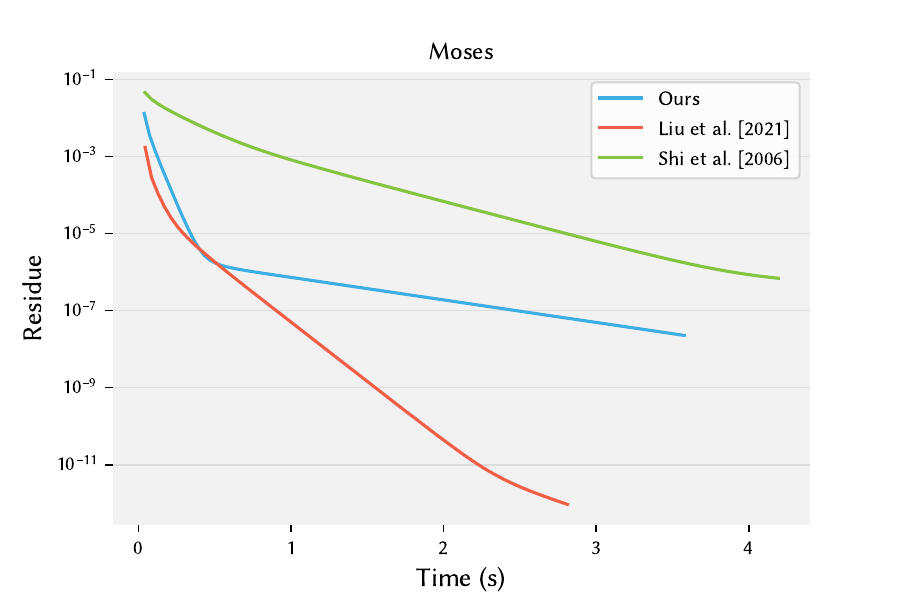}
    \end{subfigure}
    \begin{subfigure}{0.24\textwidth}
        \centering
        \includegraphics[width=\textwidth]{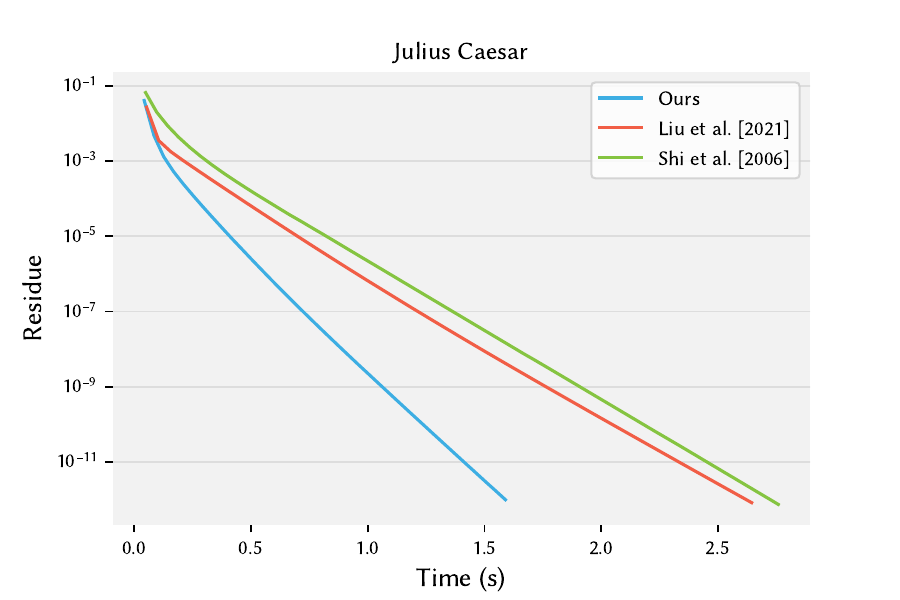}
    \end{subfigure}
    
    \begin{subfigure}{0.24\textwidth}
        \centering
        \includegraphics[width=\textwidth]{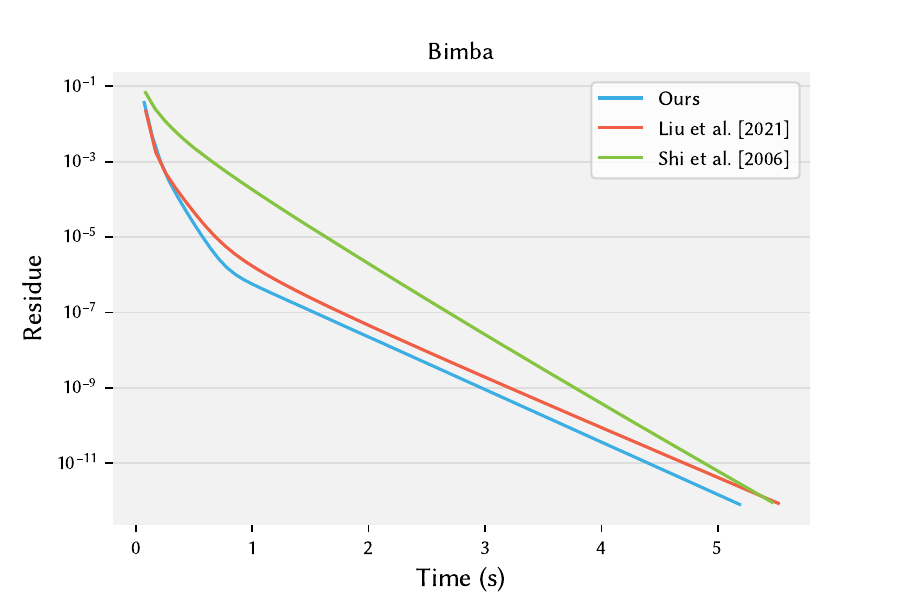}
    \end{subfigure}
    \begin{subfigure}{0.24\textwidth}
        \centering
        \includegraphics[width=\textwidth]{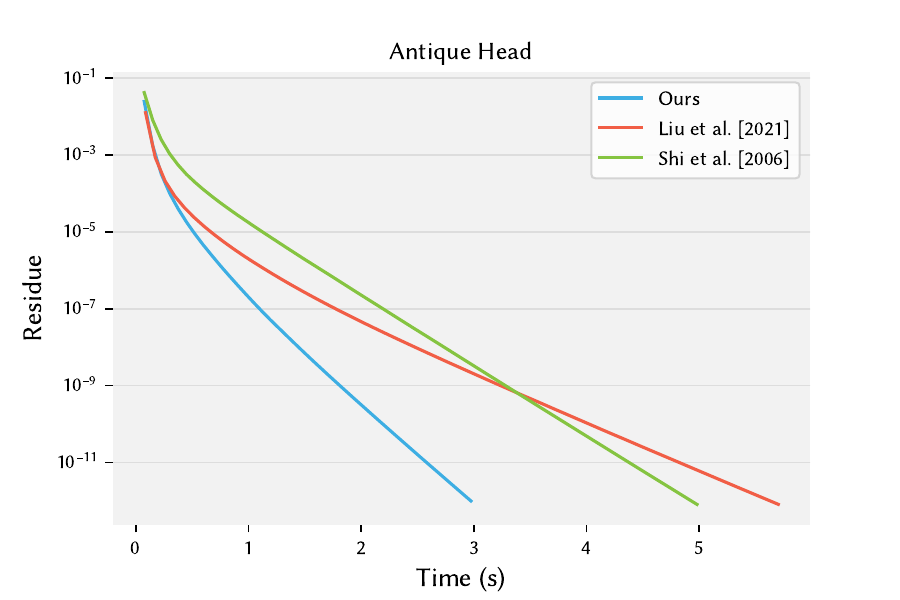}
    \end{subfigure}
    \begin{subfigure}{0.24\textwidth}
        \centering
        \includegraphics[width=\textwidth]{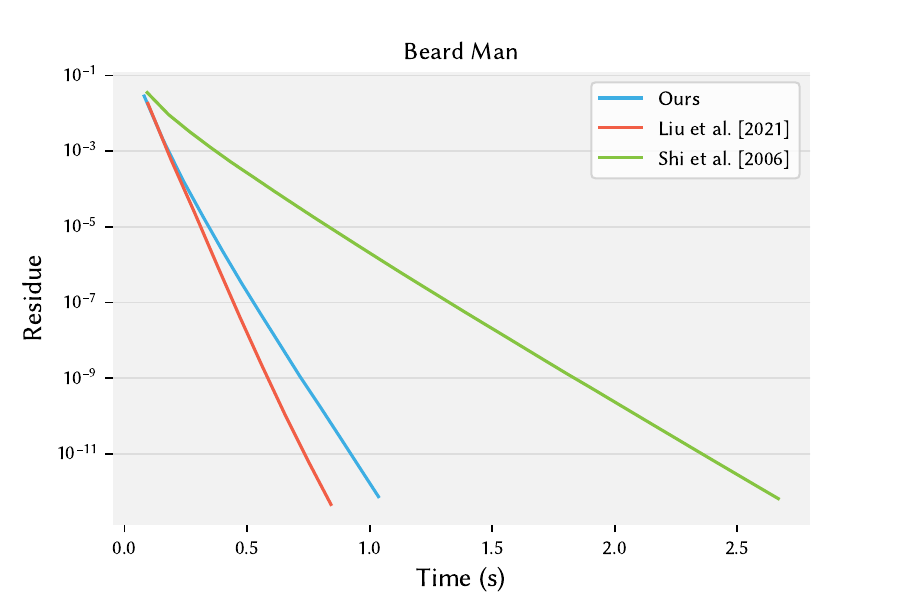}
    \end{subfigure}
    \begin{subfigure}{0.24\textwidth}
        \centering
        \includegraphics[width=\textwidth]{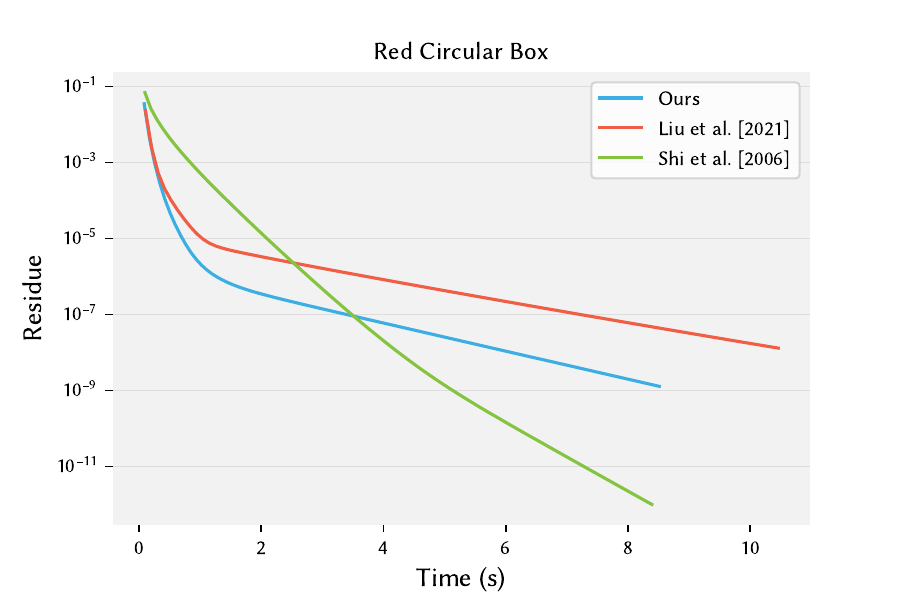}
    \end{subfigure}
    
    \begin{subfigure}{0.24\textwidth}
        \centering
        \includegraphics[width=\textwidth]{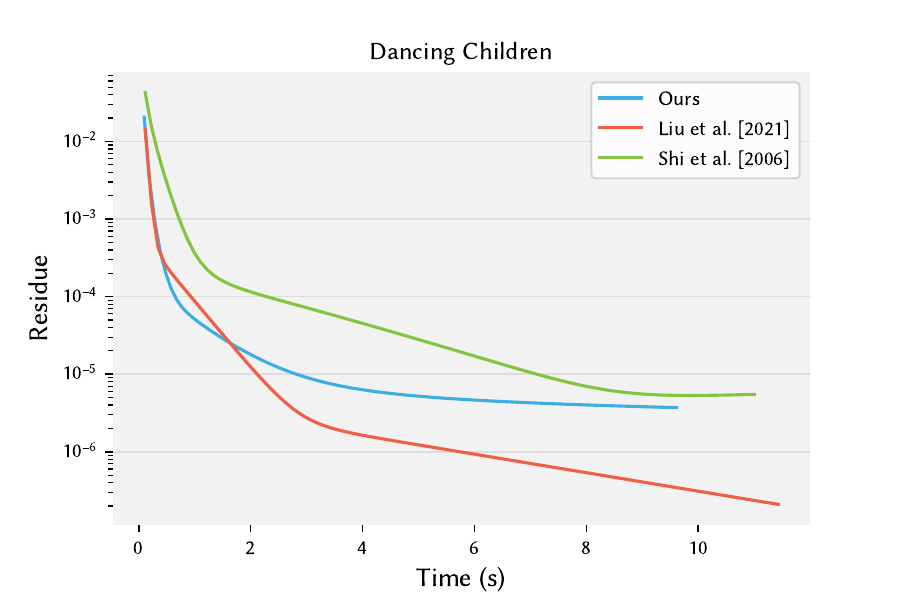}
    \end{subfigure}
    \begin{subfigure}{0.24\textwidth}
        \centering
        \includegraphics[width=\textwidth]{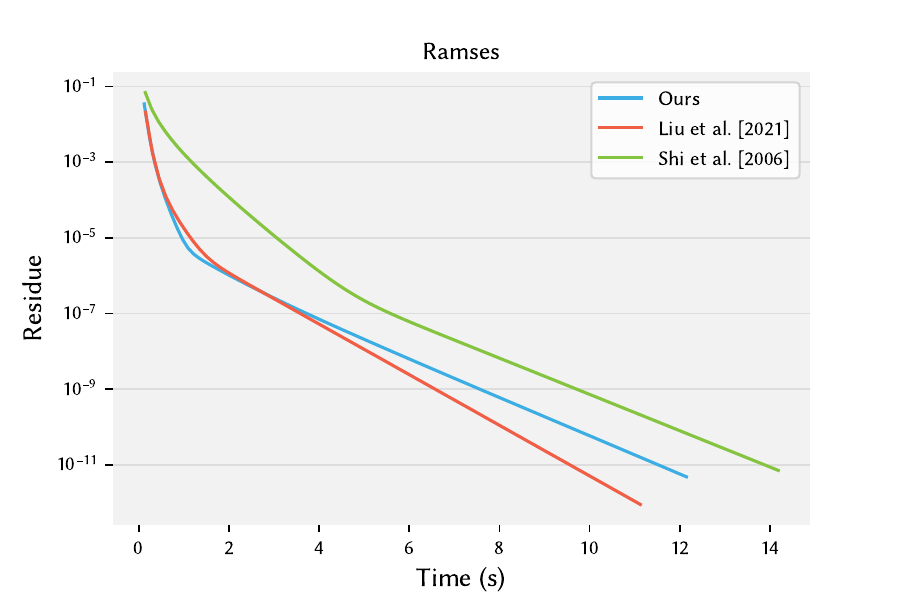}
    \end{subfigure}
    \begin{subfigure}{0.24\textwidth}
        \centering
        \includegraphics[width=\textwidth]{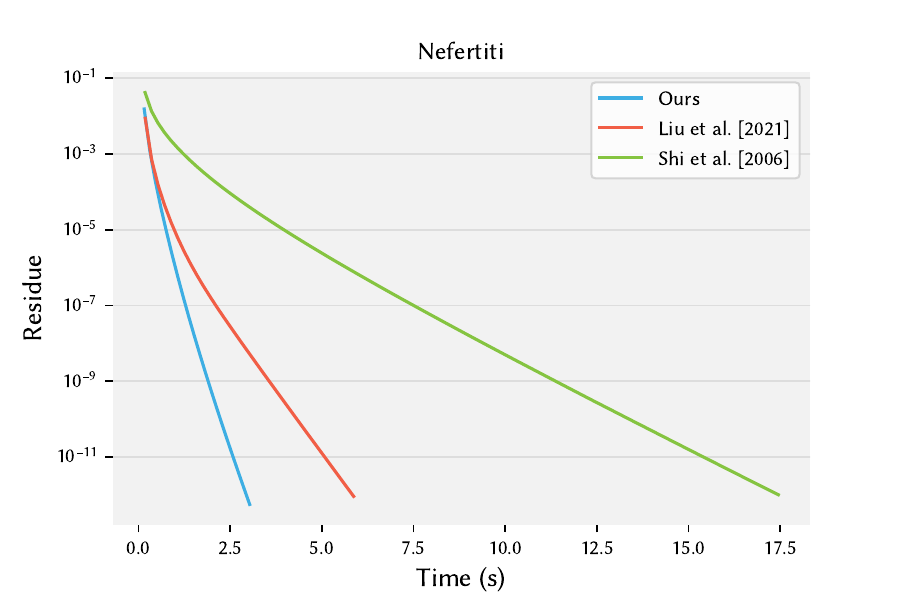}
    \end{subfigure}
    \begin{subfigure}{0.24\textwidth}
        \centering
        \includegraphics[width=\textwidth]{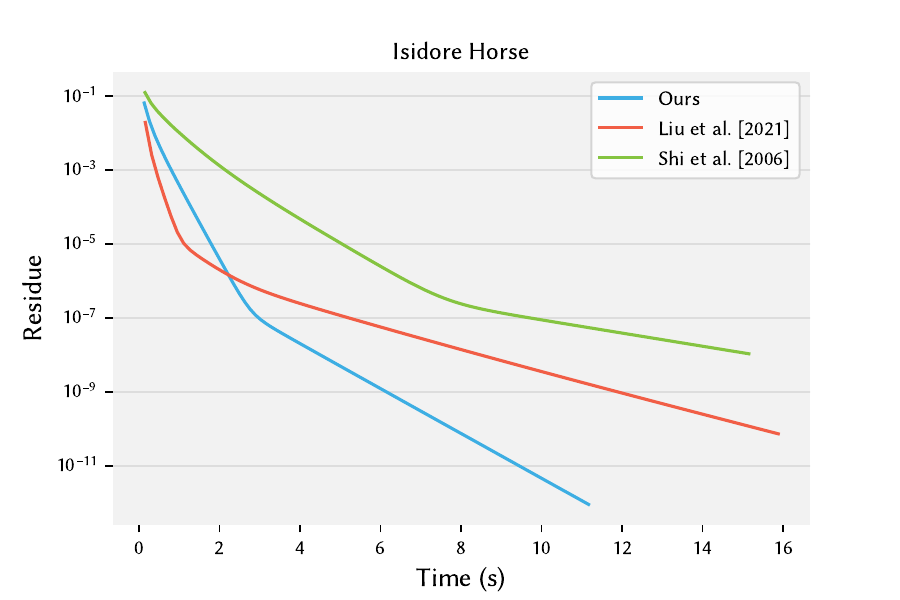}
    \end{subfigure}
    
    \begin{subfigure}{0.24\textwidth}
        \centering
        \includegraphics[width=\textwidth]{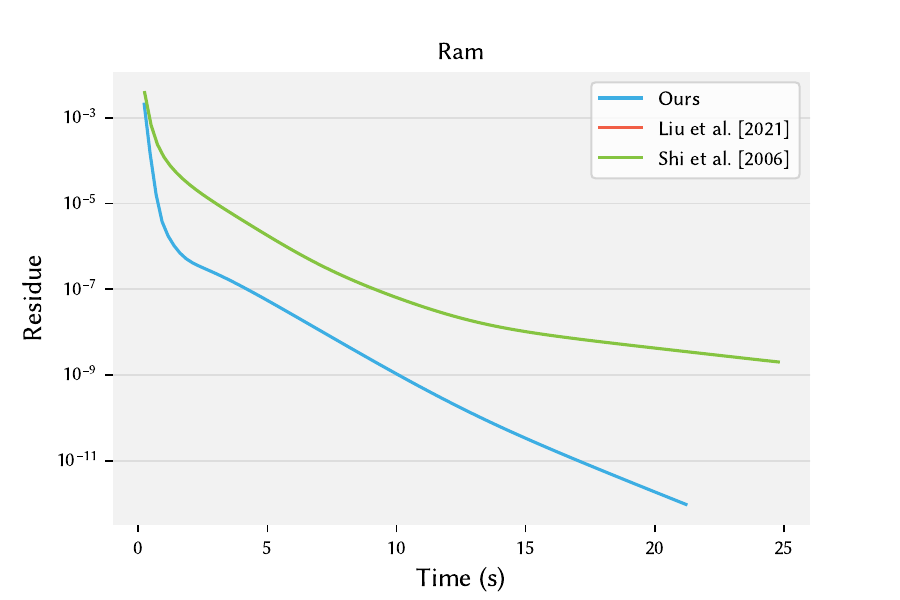}
    \end{subfigure}
    \begin{subfigure}{0.24\textwidth}
        \centering
        \includegraphics[width=\textwidth]{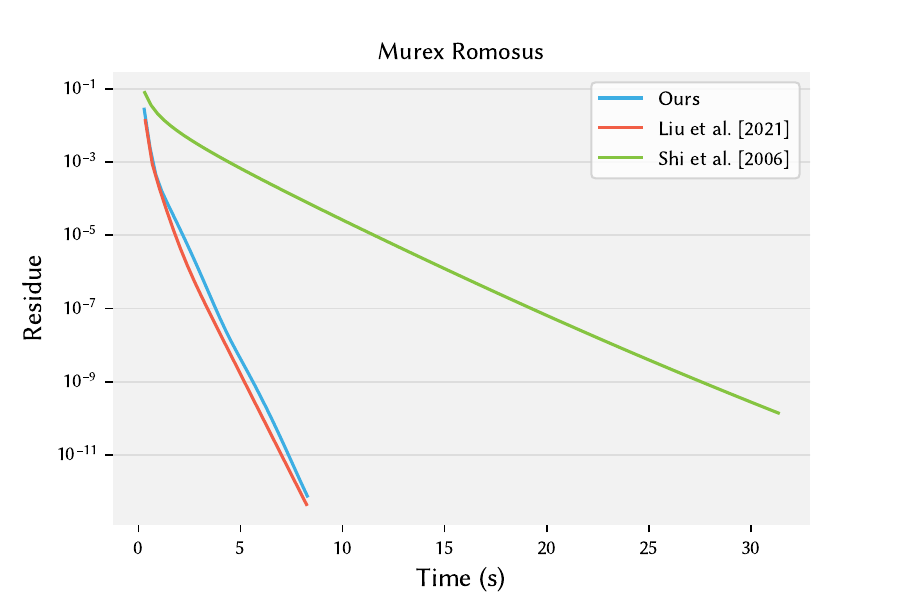}
    \end{subfigure}
    \begin{subfigure}{0.24\textwidth}
        \centering
        \includegraphics[width=\textwidth]{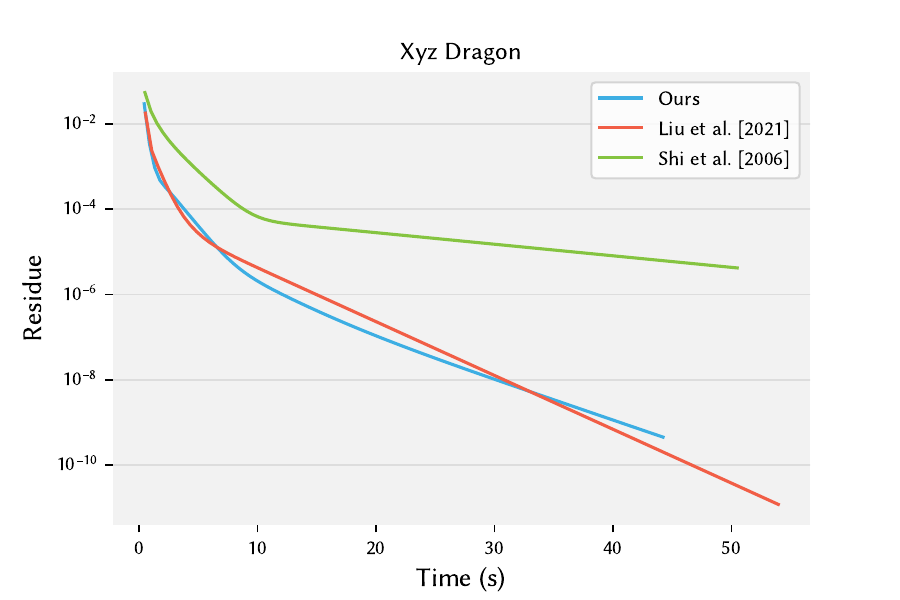}
    \end{subfigure}
    \caption{Convergence plots showing \textbf{time} on the x-axis for smoothing with $\alpha=$\num{1e-3}.}
    \label{fig:convergence}
\end{figure*}

\begin{figure*}
    \begin{subfigure}{0.24\textwidth}
        \centering
        \includegraphics[width=\textwidth]{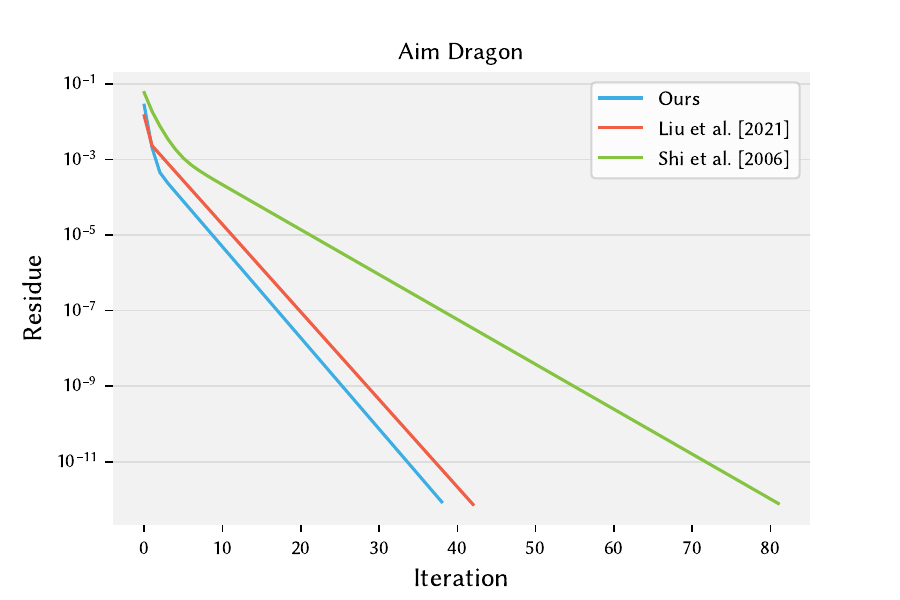}
    \end{subfigure}
    \begin{subfigure}{0.24\textwidth}
        \centering
        \includegraphics[width=\textwidth]{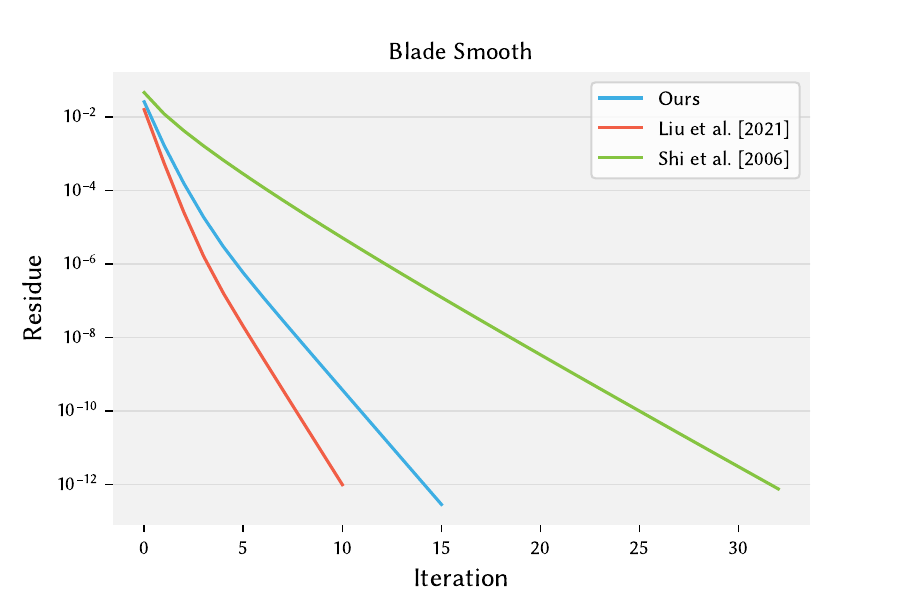}
    \end{subfigure}
    \begin{subfigure}{0.24\textwidth}
        \centering
        \includegraphics[width=\textwidth]{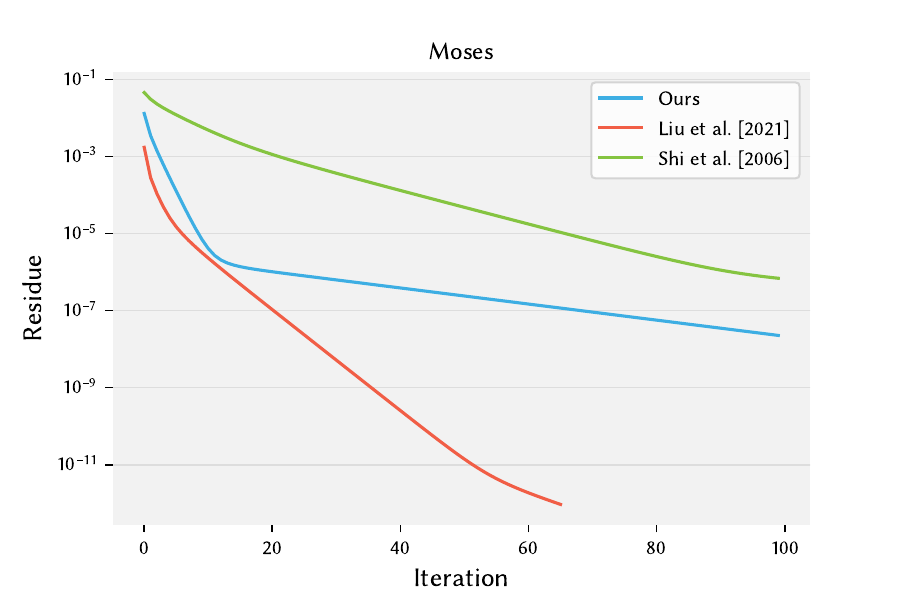}
    \end{subfigure}
    \begin{subfigure}{0.24\textwidth}
        \centering
        \includegraphics[width=\textwidth]{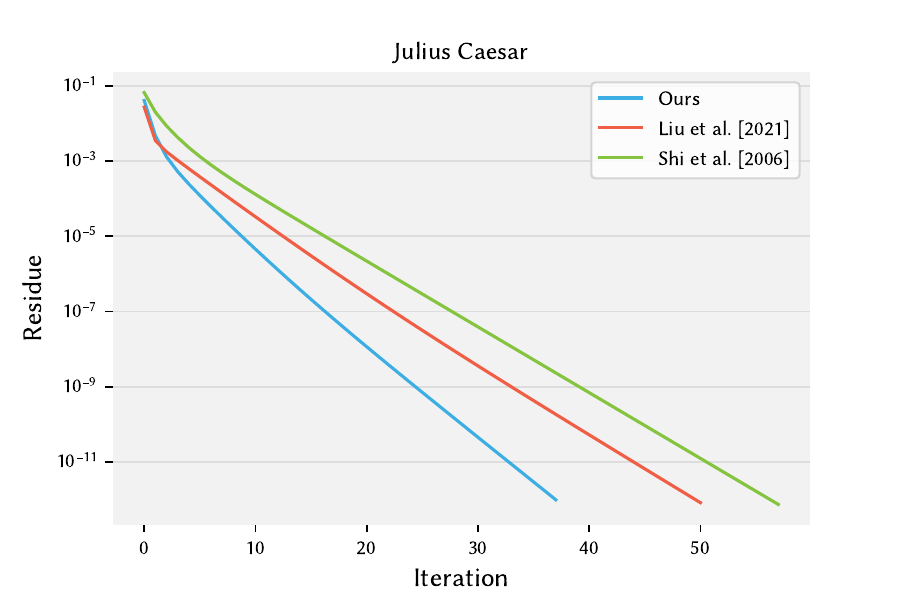}
    \end{subfigure}
    
    \begin{subfigure}{0.24\textwidth}
        \centering
        \includegraphics[width=\textwidth]{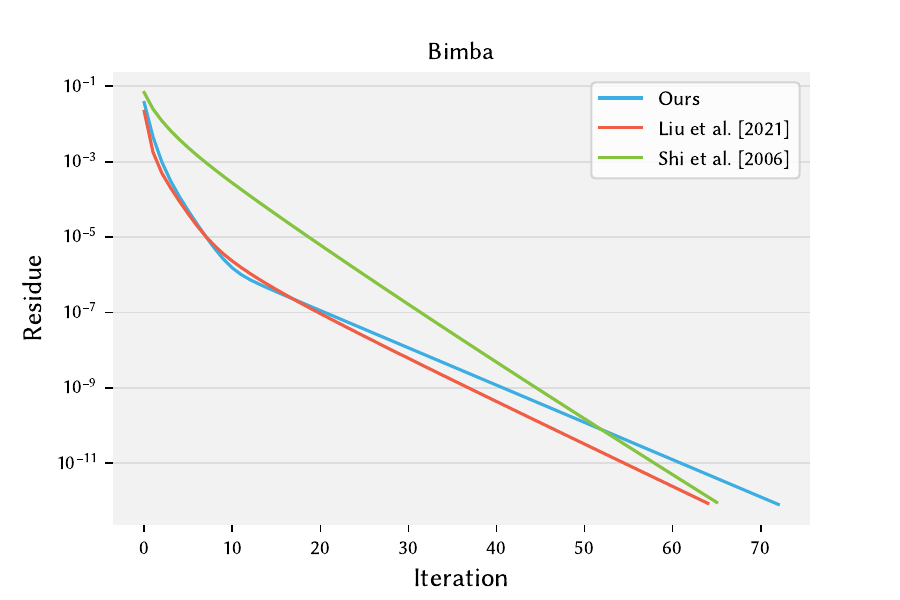}
    \end{subfigure}
    \begin{subfigure}{0.24\textwidth}
        \centering
        \includegraphics[width=\textwidth]{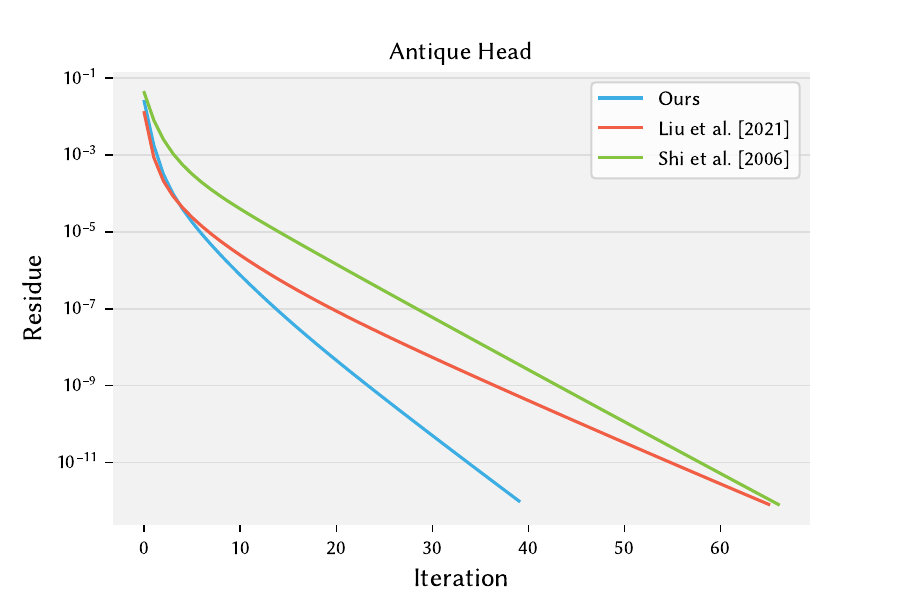}
    \end{subfigure}
    \begin{subfigure}{0.24\textwidth}
        \centering
        \includegraphics[width=\textwidth]{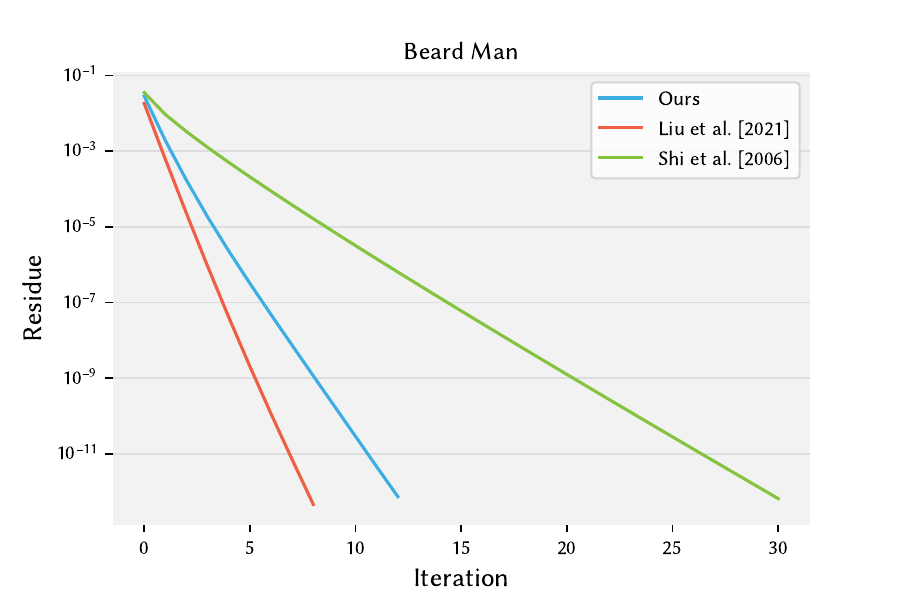}
    \end{subfigure}
    \begin{subfigure}{0.24\textwidth}
        \centering
        \includegraphics[width=\textwidth]{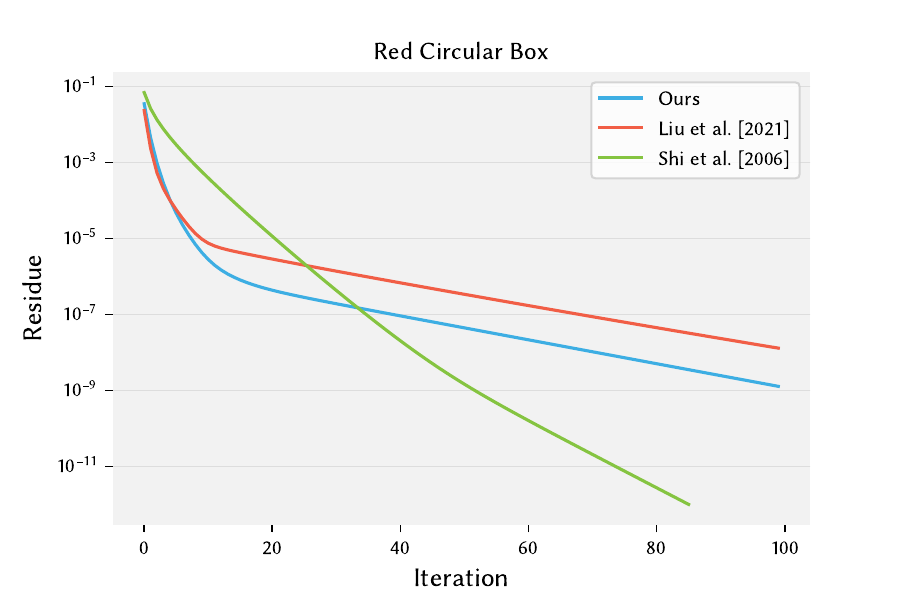}
    \end{subfigure}
    
    \begin{subfigure}{0.24\textwidth}
        \centering
        \includegraphics[width=\textwidth]{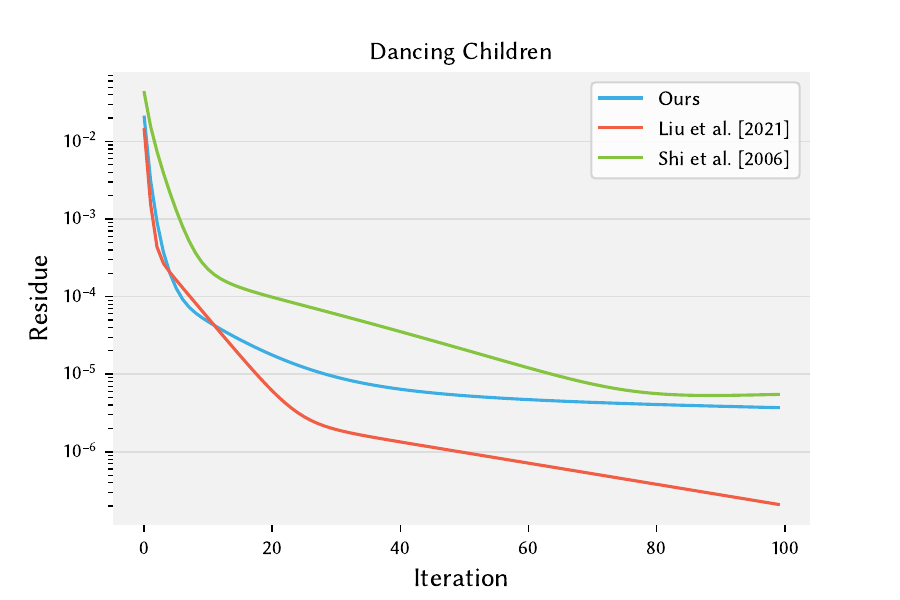}
    \end{subfigure}
    \begin{subfigure}{0.24\textwidth}
        \centering
        \includegraphics[width=\textwidth]{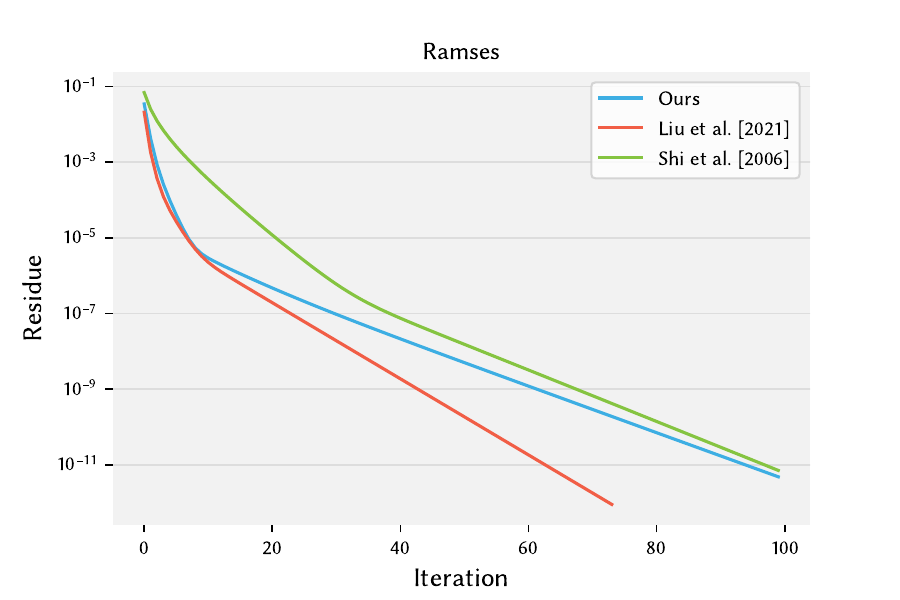}
    \end{subfigure}
    \begin{subfigure}{0.24\textwidth}
        \centering
        \includegraphics[width=\textwidth]{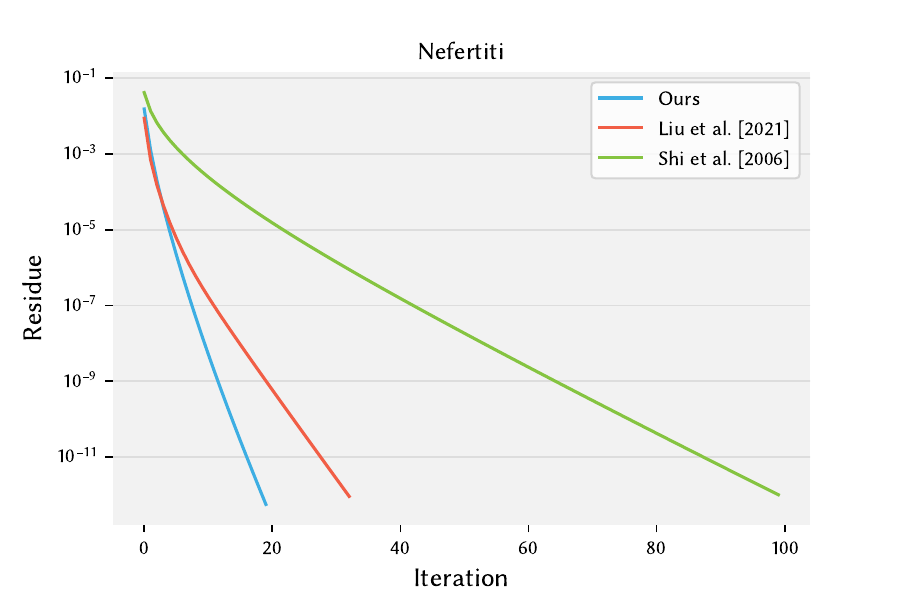}
    \end{subfigure}
    \begin{subfigure}{0.24\textwidth}
        \centering
        \includegraphics[width=\textwidth]{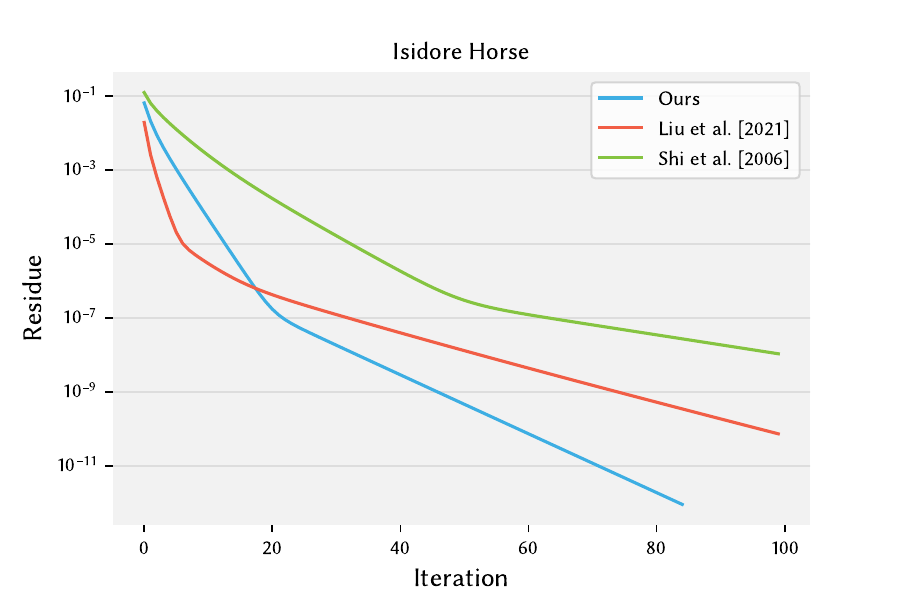}
    \end{subfigure}
    
    \begin{subfigure}{0.24\textwidth}
        \centering
        \includegraphics[width=\textwidth]{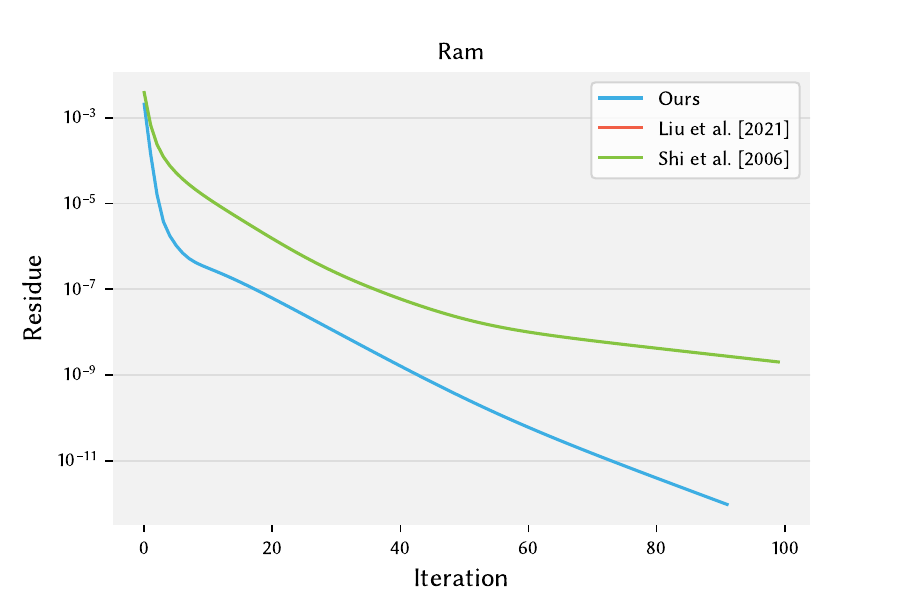}
    \end{subfigure}
    \begin{subfigure}{0.24\textwidth}
        \centering
        \includegraphics[width=\textwidth]{images/convergence/smoothing_it/murex_romosus.pdf}
    \end{subfigure}
    \begin{subfigure}{0.24\textwidth}
        \centering
        \includegraphics[width=\textwidth]{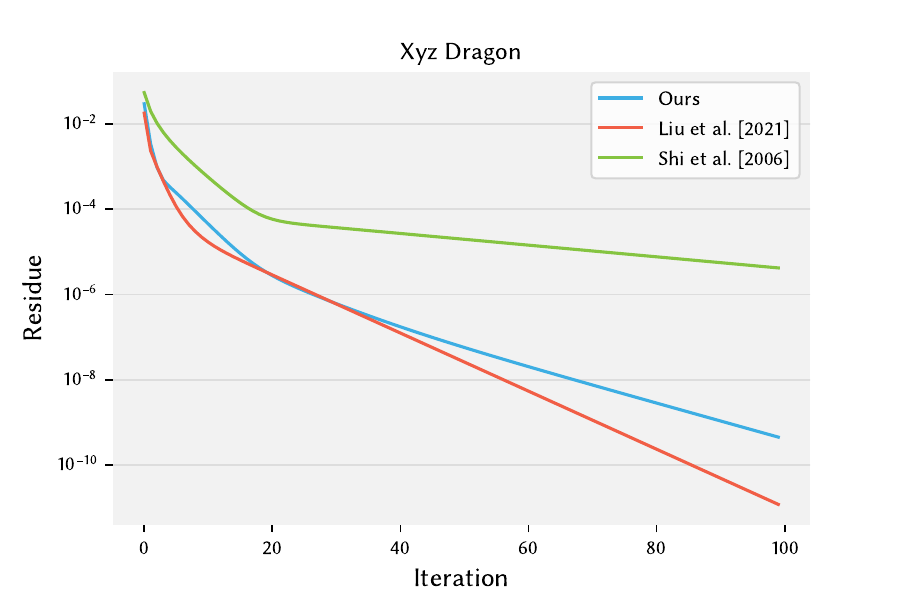}
    \end{subfigure}
    \caption{Convergence plots showing \textbf{iterations} on the x-axis for smoothing with $\alpha=$\num{1e-3}.}
    \label{fig:convergence_iterations}
\end{figure*}

We also report a comparison with a data smoothing problem with $\alpha=$\num{1e-3} for manifold meshes in \autoref{tab:comparisonall} and for non-manifold meshes and point clouds in \autoref{tab:comparison_nm_pc}. Convergence plots for data smoothing on the manifold meshes in the main paper are shown in \autoref{fig:convergence} and \autoref{fig:convergence_iterations}.

\begin{table*}[h]
\caption{Comparison of our hierarchy construction and solver for a \textbf{Poisson problem} with $\eta=$\num{1e-6} mass matrix coefficient and tolerance of \num{1e-4}. The maximum number of iterations for iterative solvers is set to 100.}
\centering
\resizebox{\textwidth}{!}{
\begin{tabular}{@{}lr|rrr|rrr|rrr||rrr|rrr||rr|rr@{}}
\hline

 \multicolumn{1}{c}{\multirow{2}{*}{\textbf{Model}}} & \multicolumn{1}{c|}{\multirow{2}{*}{\textbf{\#Vert}}} & \multicolumn{3}{c|}{\textbf{\ourmethod~(Ours)}} & \multicolumn{3}{c|}{\textbf{\citet{Liu2021}}} & \multicolumn{3}{c||}{\textbf{\citet{Shi2006}}} & \multicolumn{3}{c|}{\textbf{AMG-RS}} & \multicolumn{3}{c||}{\textbf{AMG-SA}} & \multicolumn{2}{c|}{\textbf{Eigen}} & \multicolumn{2}{c}{\textbf{\textsc{Pardiso}}}  \\
\cline{3-21}
\multicolumn{1}{c}{} & \multicolumn{1}{c|}{} & \small{Hier} & \small{\#It} & \small{Solve} & \small{Hier} & \small{\#It} & \small{Solve} & \small{Hier} & \small{\#It} & \small{Solve} & \small{Hier} & \small{\#It} & \small{Solve} & \small{Hier} & \small{\#It} & \small{Solve} & \small{Fact.} & \small{Subst.} & \small{Fact.} & \small{Subst.}   \\
\hline
 Beetle           & 19k   &        0.01 &       28 &        0.06 &              0.55 &       18 &        0.05 &              0.01 &       33 &        0.09 &           0.03 &      100 &        0.29 &           0.03 &      100 &        0.20 &             0.02 &              0.00 &             0.06 &              0.00 \\
 Ogre             & 19k   &        0.02 &        6 &        0.02 &              0.54 &        7 &        0.03 &              0.01 &       13 &        0.04 &           0.03 &      100 &        0.27 &           0.04 &       40 &        0.08 &             0.02 &              0.00 &             0.06 &              0.00 \\
 Screwdriver      & 27k   &        0.02 &        4 &        0.03 &              0.75 &        4 &        0.03 &              0.01 &        9 &        0.04 &           0.03 &      100 &        0.35 &           0.05 &       24 &        0.07 &             0.05 &              0.00 &             0.09 &              0.00 \\
 Mumble           & 34k   &        0.03 &       10 &        0.05 &              0.97 &        6 &        0.04 &              0.02 &        9 &        0.04 &           0.04 &      100 &        0.41 &           0.05 &       57 &        0.17 &             0.04 &              0.00 &             0.11 &              0.01 \\
 Horse            & 48k   &        0.04 &        7 &        0.05 &              1.42 &        6 &        0.06 &              0.02 &       12 &        0.08 &           0.05 &      100 &        0.58 &           0.08 &       41 &        0.19 &             0.10 &              0.00 &             0.14 &              0.01 \\
 Laurent's Hand     & 50k   &        0.04 &        5 &        0.07 &              1.59 &        4 &        0.06 &              0.02 &       23 &        0.18 &           0.06 &      100 &        0.66 &           0.10 &       36 &        0.21 &             0.10 &              0.00 &             0.18 &              0.01 \\
 Dinosaur         & 56k   &        0.04 &        7 &        0.06 &              1.71 &        5 &        0.06 &              0.02 &       17 &        0.14 &           0.06 &      100 &        0.65 &           0.09 &       49 &        0.28 &             0.08 &              0.00 &             0.18 &              0.01 \\
 Heart            & 78k   &        0.06 &       15 &        0.18 &              2.90 &       29 &        0.38 &              0.04 &       33 &        0.43 &           0.08 &      100 &        1.06 &           0.13 &      100 &        0.95 &             0.17 &              0.01 &             0.27 &              0.01 \\
 Hannya Mask      & 83k   &        0.07 &       15 &        0.19 &              2.56 &        4 &        0.09 &              0.05 &       19 &        0.26 &           0.09 &      100 &        1.03 &           0.13 &      100 &        0.91 &             0.17 &              0.01 &             0.30 &              0.02 \\
 Trex             & 100k  &        0.10 &        9 &        0.16 &              3.40 &        4 &        0.14 &              0.05 &       42 &        0.64 &           0.12 &      100 &        1.39 &           0.22 &       97 &        1.10 &             0.18 &              0.01 &             0.38 &              0.02 \\
 The Thinker      & 110k  &        0.09 &        4 &        0.08 &              3.50 &        4 &        0.11 &              0.04 &       13 &        0.19 &           0.10 &      100 &        1.10 &           0.20 &       26 &        0.26 &             0.36 &              0.01 &             0.39 &              0.02 \\
 Egea            & 134k  &        0.11 &        4 &        0.13 &              4.42 &        4 &        0.14 &              0.06 &       15 &        0.31 &           0.16 &      100 &        1.79 &           0.23 &       27 &        0.36 &             0.94 &              0.02 &             0.50 &              0.03 \\
 Sappho's Head          & 140k  &        0.11 &        6 &        0.17 &              4.36 &        7 &        0.21 &              0.07 &       14 &        0.35 &           0.15 &      100 &        1.79 &           0.24 &       45 &        0.73 &             0.38 &              0.01 &             0.51 &              0.03 \\
 Human Torso      & 142k  &        0.15 &        4 &        0.15 &              4.97 &        5 &        0.20 &              0.06 &       40 &        0.85 &           0.16 &      100 &        1.86 &           0.25 &       44 &        0.72 &             0.73 &              0.02 &             0.57 &              0.03 \\
 Aim Dragon       & 152k  &        0.14 &        7 &        0.19 &              5.14 &        8 &        0.27 &              0.06 &       27 &        0.62 &           0.15 &       26 &        0.46 &           0.31 &       29 &        0.48 &             0.81 &              0.02 &             0.58 &              0.04 \\
 Armadillo        & 172k  &        0.16 &        9 &        0.28 &              5.97 &        7 &        0.29 &              0.08 &       27 &        0.76 &           0.20 &      100 &        2.37 &           0.36 &       52 &        1.04 &             0.54 &              0.02 &             0.62 &              0.04 \\
 Ronaldo          & 176k  &        0.18 &        8 &        0.28 &              6.32 &        6 &        0.30 &              0.10 &       40 &        1.19 &           0.21 &      100 &        2.74 &           0.40 &       75 &        1.72 &             0.60 &              0.02 &             0.69 &              0.04 \\
 Isis             & 187k  &        0.16 &        4 &        0.18 &              6.25 &        3 &        0.17 &              0.09 &       15 &        0.45 &           0.22 &      100 &        2.66 &           0.31 &       70 &        1.39 &             1.12 &              0.02 &             0.65 &              0.04 \\
 Blade Smooth     & 195k  &        0.15 &        4 &        0.17 &              5.87 &        3 &        0.17 &              0.09 &       18 &        0.60 &           0.18 &      100 &        2.42 &           0.40 &       42 &        0.94 &             0.76 &              0.02 &             0.69 &              0.04 \\
 Max Planck       & 199k  &        0.16 &        7 &        0.26 &              5.98 &        4 &        0.19 &              0.08 &       19 &        0.55 &           0.17 &      100 &        2.20 &           0.30 &       34 &        0.69 &             1.94 &              0.03 &             0.67 &              0.05 \\
 Vase-Lion        & 200k  &        0.17 &        4 &        0.20 &              6.54 &        4 &        0.22 &              0.12 &       10 &        0.43 &           0.21 &      100 &        2.68 &           0.33 &       36 &        0.88 &             0.40 &              0.02 &             0.79 &              0.04 \\
 Duck             & 204k  &        0.16 &        8 &        0.26 &              5.91 &        7 &        0.27 &              0.08 &       16 &        0.45 &           0.18 &      100 &        2.47 &           0.37 &       69 &        1.39 &             1.68 &              0.03 &             0.64 &              0.05 \\
 Mouse            & 214k  &        0.19 &        6 &        0.23 &              6.33 &        4 &        0.20 &              0.10 &       21 &        0.61 &           0.20 &      100 &        2.60 &           0.39 &       60 &        1.28 &             1.65 &              0.03 &             0.70 &              0.05 \\
 Wolf Skull       & 228k  &        0.23 &        7 &        0.33 &              8.07 &        5 &        0.33 &              0.13 &       38 &        1.52 &           0.27 &      100 &        3.47 &           0.53 &       53 &        1.59 &             0.39 &              0.02 &             0.92 &              0.05 \\
 Moses            & 258k  &        0.42 &       12 &        0.55 &              8.72 &        5 &        0.35 &              0.30 &      100 &        4.30 &           0.27 &      100 &        3.64 &           0.55 &      100 &        3.33 &             0.90 &              0.02 &             1.04 &              0.05 \\
 Rockerarm        & 271k  &        0.27 &        3 &        0.20 &              9.20 &        3 &        0.27 &              0.16 &        5 &        0.34 &           0.28 &      100 &        2.90 &           0.52 &       22 &        0.53 &             1.79 &              0.03 &             1.01 &              0.06 \\
 Pulley2          & 293k  &        0.24 &        5 &        0.32 &              8.87 &        4 &        0.30 &              0.14 &       24 &        1.04 &           0.28 &      100 &        3.67 &           0.58 &       38 &        1.19 &             3.03 &              0.04 &             1.07 &              0.07 \\
 Heraklion        & 350k  &        0.36 &       20 &        1.18 &             12.95 &        7 &        0.67 &              0.21 &      100 &        5.79 &           0.46 &      100 &        5.82 &           0.80 &      100 &        4.54 &             2.39 &              0.04 &             1.45 &              0.08 \\
 Julius Caesar    & 387k  &        0.30 &       11 &        0.58 &             12.10 &       17 &        1.09 &              0.16 &       28 &        1.54 &           0.39 &      100 &        4.89 &           0.77 &       70 &        2.79 &             5.07 &              0.06 &             1.29 &              0.09 \\
 Goyle            & 393k  &        0.33 &        2 &        0.21 &             11.44 &        2 &        0.26 &              0.21 &       12 &        0.73 &           0.39 &      100 &        4.75 &           0.79 &       25 &        0.97 &             8.43 &              0.07 &             1.32 &              0.10 \\
 Eros             & 476k  &        0.41 &        NaN &        NaN &             14.78 &        5 &        0.55 &              0.21 &       28 &        1.90 &           0.46 &      100 &        6.01 &           0.97 &       71 &        3.55 &             4.03 &              0.06 &             1.76 &              0.11 \\
 Roal             & 484k  &        0.47 &        5 &        0.54 &             16.52 &        4 &        0.59 &              0.37 &       37 &        3.27 &           0.53 &      100 &        6.86 &           0.84 &       53 &        3.08 &             1.34 &              0.04 &             2.08 &              0.09 \\
 Skeleton         & 494k  &        0.71 &        9 &        1.10 &             20.60 &        NaN &        NaN &              0.35 &      100 &        9.18 &           0.66 &      100 &        9.42 &           1.63 &      100 &        7.82 &             3.33 &              0.05 &             2.33 &              0.11 \\
 Bimba            & 502k  &        0.43 &        7 &        0.65 &             15.58 &        6 &        0.74 &              0.24 &       48 &        4.16 &           0.51 &      100 &        6.97 &           1.15 &       69 &        4.45 &             3.54 &              0.05 &             2.13 &              0.10 \\
 Oil Pump         & 570k  &        0.47 &       19 &        1.57 &             18.76 &       12 &        1.31 &              0.25 &       45 &        3.93 &           0.55 &      100 &        7.29 &           1.27 &       74 &        4.92 &             6.85 &              0.07 &             2.30 &              0.13 \\
 Antique Head     & 651k  &        0.53 &        4 &        0.51 &             20.07 &        4 &        0.60 &              0.28 &       14 &        1.22 &           0.64 &      100 &        8.48 &           1.37 &       67 &        4.33 &            15.02 &              0.11 &             2.43 &              0.17 \\
 Pulley           & 660k  &        0.54 &       19 &        1.81 &             21.54 &       16 &        1.88 &              0.28 &       53 &        5.23 &           0.63 &      100 &        8.46 &           1.49 &       58 &        4.31 &            12.94 &              0.10 &             2.68 &              0.16 \\
 Beard Man        & 691k  &        0.59 &        4 &        0.56 &             22.15 &        3 &        0.58 &              0.26 &       14 &        1.43 &           0.52 &      100 &        7.07 &           1.46 &       14 &        0.96 &            24.57 &              0.14 &             2.72 &              0.19 \\
 Red Circular Box & 701k  &        0.64 &        6 &        0.74 &             22.97 &        6 &        0.97 &              0.34 &       66 &        6.67 &           0.72 &      100 &        8.98 &           1.51 &       66 &        5.01 &            17.76 &              0.11 &             2.84 &              0.17 \\
 Dancing Children & 724k  &        0.69 &        9 &        1.10 &             23.21 &        8 &        1.25 &              0.41 &       39 &        4.54 &           0.68 &      100 &        9.32 &           1.23 &      100 &        8.70 &             6.18 &              0.09 &             2.78 &              0.17 \\
 John The Baptist & 750k  &        0.73 &       69 &        7.74 &             29.90 &       10 &        1.76 &              0.47 &       81 &       10.56 &           0.86 &      100 &       10.53 &           1.86 &      100 &        9.77 &             6.43 &              0.08 &             3.08 &              0.17 \\
 Ramses           & 826k  &        0.79 &        7 &        1.21 &             28.90 &        5 &        1.18 &              0.47 &       40 &        6.03 &           0.91 &      100 &       11.82 &           2.08 &       49 &        5.40 &             6.24 &              0.09 &             3.60 &              0.18 \\
 Nicolo Da Uzzano & 946k  &        0.76 &        7 &        1.02 &             29.88 &        6 &        1.16 &              0.40 &       33 &        4.15 &           0.85 &      100 &       12.08 &           2.05 &       82 &        8.06 &            17.75 &              0.16 &             3.53 &              0.24 \\
 Raptor           & 1m &        0.89 &        NaN &        NaN &             32.27 &        NaN &        NaN &              0.43 &        NaN &        NaN &           1.07 &      100 &       13.98 &           1.83 &      100 &       12.14 &             0.79 &              0.00 &             3.77 &              0.20 \\
 Nefertiti        & 1m &        0.89 &        4 &        0.94 &             34.22 &        4 &        1.18 &              0.56 &       65 &       11.69 &           1.09 &      100 &       14.74 &           2.58 &       46 &        6.14 &             9.15 &              0.10 &             4.54 &              0.22 \\
 Isidore Horse    & 1.1m &        1.12 &       11 &        1.90 &             35.60 &        5 &        1.27 &              0.50 &       70 &       11.03 &           1.11 &      100 &       14.60 &           2.50 &       88 &       10.72 &            24.01 &              0.17 &             4.56 &              0.28 \\
 Horse Head       & 1.3m &        2.42 &       44 &       10.76 &             54.11 &       10 &        4.52 &              1.00 &      100 &       24.52 &           2.12 &      100 &       24.57 &           4.25 &      100 &       20.97 &            12.99 &              0.13 &             6.55 &              0.30 \\
 Ram              & 1.3m &        2.40 &        3 &        1.95 &             56.18 &        NaN &        NaN &              1.14 &       49 &       13.39 &           2.45 &      100 &       25.95 &           4.72 &      100 &       21.71 &            17.07 &              0.15 &             6.99 &              0.33 \\
 Murex Romosus    & 1.8m &        2.32 &        6 &        2.85 &             73.05 &        5 &        3.32 &              0.99 &       63 &       20.79 &           2.38 &      100 &       29.83 &           5.08 &       58 &       15.28 &            40.06 &              0.26 &             9.13 &              0.44 \\
 XYZ Dragon       & 3.6m &        3.24 &        9 &        5.32 &            121.97 &        7 &        5.32 &              1.57 &       55 &       28.95 &           3.16 &      100 &       43.75 &           9.14 &       75 &       30.54 &            77.62 &              0.69 &            15.88 &              0.94 \\
\hline
\end{tabular}
}
\label{tab:comparisonallpoisson}
\end{table*}

\begin{table*}[h]
\caption{Comparison of our hierarchy construction and solver for \textbf{data smoothing} of a random function with smoothing coefficient $\alpha=$\num{1e-3} and tolerance of \num{1e-4}. The maximum number of iterations for iterative solvers is set to 100.}
\centering
\resizebox{\textwidth}{!}{
\begin{tabular}{@{}lr|rrr|rrr|rrr||rrr|rrr||rr|rr@{}}
\hline

 \multicolumn{1}{c}{\multirow{2}{*}{\textbf{Model}}} & \multicolumn{1}{c|}{\multirow{2}{*}{\textbf{\#Vert}}} & \multicolumn{3}{c|}{\textbf{\ourmethod~(Ours)}} & \multicolumn{3}{c|}{\textbf{\citet{Liu2021}}} & \multicolumn{3}{c||}{\textbf{\citet{Shi2006}}} & \multicolumn{3}{c|}{\textbf{AMG-RS}} & \multicolumn{3}{c||}{\textbf{AMG-SA}} & \multicolumn{2}{c|}{\textbf{Eigen}} & \multicolumn{2}{c}{\textbf{\textsc{Pardiso}}}  \\
\cline{3-21}
\multicolumn{1}{c}{} & \multicolumn{1}{c|}{} & \small{Hier} & \small{\#It} & \small{Solve} & \small{Hier} & \small{\#It} & \small{Solve} & \small{Hier} & \small{\#It} & \small{Solve} & \small{Hier} & \small{\#It} & \small{Solve} & \small{Hier} & \small{\#It} & \small{Solve} & \small{Fact.} & \small{Subst.} & \small{Fact.} & \small{Subst.}   \\
\hline
  Beetle           & 19k   &        0.01 &       27 &        0.06 &              0.55 &       12 &        0.04 &              0.01 &       24 &        0.07 &           0.03 &       75 &        0.22 &           0.04 &       92 &        0.18 &             0.02 &              0.00 &             0.06 &              0.00 \\
 Ogre             & 19k   &        0.01 &        6 &        0.02 &              0.54 &        7 &        0.03 &              0.01 &        8 &        0.02 &           0.03 &       12 &        0.03 &           0.04 &       16 &        0.03 &             0.02 &              0.00 &             0.06 &              0.00 \\
 Screwdriver      & 27k   &        0.02 &        4 &        0.03 &              0.75 &        4 &        0.03 &              0.01 &        7 &        0.03 &           0.03 &       14 &        0.05 &           0.04 &       12 &        0.03 &             0.05 &              0.00 &             0.08 &              0.00 \\
 Mumble           & 34k   &        0.03 &        9 &        0.04 &              1.00 &        6 &        0.04 &              0.02 &        7 &        0.04 &           0.04 &       29 &        0.12 &           0.05 &       46 &        0.13 &             0.04 &              0.00 &             0.11 &              0.01 \\
 Horse            & 48k   &        0.04 &        6 &        0.04 &              1.38 &        6 &        0.06 &              0.02 &        8 &        0.06 &           0.05 &       34 &        0.20 &           0.07 &       15 &        0.07 &             0.10 &              0.00 &             0.14 &              0.01 \\
 Laurent's Hand     & 50k   &        0.04 &        5 &        0.07 &              1.56 &        4 &        0.06 &              0.02 &        9 &        0.08 &           0.06 &       17 &        0.11 &           0.11 &       12 &        0.07 &             0.10 &              0.00 &             0.17 &              0.01 \\
 Dinosaur         & 56k   &        0.04 &        5 &        0.05 &              1.72 &        4 &        0.06 &              0.02 &        7 &        0.07 &           0.06 &       57 &        0.37 &           0.09 &       22 &        0.13 &             0.08 &              0.00 &             0.18 &              0.01 \\
 Heart            & 78k   &        0.06 &       16 &        0.19 &              2.92 &       26 &        0.34 &              0.04 &       16 &        0.22 &           0.08 &        7 &        0.08 &           0.13 &       60 &        0.58 &             0.17 &              0.01 &             0.27 &              0.01 \\
 Hannya Mask      & 83k   &        0.07 &        8 &        0.11 &              2.56 &        3 &        0.08 &              0.05 &       11 &        0.16 &           0.08 &       34 &        0.35 &           0.13 &      100 &        0.92 &             0.17 &              0.01 &             0.30 &              0.02 \\
 Trex             & 100k  &        0.10 &        7 &        0.13 &              3.40 &        4 &        0.14 &              0.04 &       12 &        0.21 &           0.13 &      100 &        1.40 &           0.18 &       23 &        0.26 &             0.20 &              0.01 &             0.38 &              0.02 \\
 The Thinker      & 110k  &        0.09 &        4 &        0.08 &              3.48 &        3 &        0.10 &              0.04 &        7 &        0.11 &           0.10 &       20 &        0.22 &           0.21 &       11 &        0.12 &             0.35 &              0.01 &             0.39 &              0.02 \\
 Egea            & 134k  &        0.11 &        5 &        0.14 &              4.40 &        4 &        0.14 &              0.06 &        7 &        0.17 &           0.15 &       35 &        0.63 &           0.22 &       12 &        0.16 &             0.90 &              0.02 &             0.50 &              0.03 \\
 Sappho's Head          & 140k  &        0.11 &        6 &        0.16 &              4.38 &        7 &        0.22 &              0.07 &        8 &        0.22 &           0.14 &      100 &        1.79 &           0.29 &       16 &        0.26 &             0.38 &              0.01 &             0.51 &              0.03 \\
 Human Torso      & 142k  &        0.15 &        4 &        0.15 &              4.88 &        4 &        0.17 &              0.06 &       10 &        0.24 &           0.16 &       97 &        1.77 &           0.25 &       12 &        0.19 &             0.74 &              0.02 &             0.57 &              0.04 \\
 Aim Dragon       & 152k  &        0.13 &        6 &        0.17 &              5.07 &        8 &        0.27 &              0.06 &       14 &        0.35 &           0.15 &        6 &        0.11 &           0.31 &       10 &        0.17 &             0.80 &              0.02 &             0.58 &              0.04 \\
 Armadillo        & 172k  &        0.15 &        9 &        0.27 &              5.88 &        7 &        0.28 &              0.08 &       20 &        0.57 &           0.20 &       15 &        0.36 &           0.35 &       34 &        0.68 &             0.54 &              0.02 &             0.62 &              0.04 \\
 Ronaldo          & 176k  &        0.18 &        7 &        0.26 &              6.35 &        5 &        0.26 &              0.10 &       13 &        0.44 &           0.20 &       64 &        1.75 &           0.40 &       38 &        0.88 &             0.62 &              0.02 &             0.69 &              0.04 \\
 Isis             & 187k  &        0.16 &        3 &        0.16 &              6.29 &        3 &        0.18 &              0.09 &       10 &        0.32 &           0.22 &       23 &        0.61 &           0.31 &       46 &        0.92 &             1.13 &              0.02 &             0.65 &              0.04 \\
 Blade Smooth     & 195k  &        0.15 &        4 &        0.17 &              5.85 &        3 &        0.17 &              0.09 &        8 &        0.30 &           0.18 &       21 &        0.51 &           0.40 &       16 &        0.36 &             0.76 &              0.02 &             0.69 &              0.04 \\
 Max Planck       & 199k  &        0.16 &        7 &        0.26 &              5.95 &        4 &        0.20 &              0.08 &        8 &        0.26 &           0.18 &      100 &        2.20 &           0.30 &       20 &        0.42 &             1.97 &              0.03 &             0.67 &              0.05 \\
 Vase-Lion        & 200k  &        0.17 &        3 &        0.17 &              6.34 &        2 &        0.15 &              0.12 &        3 &        0.17 &           0.20 &       12 &        0.32 &           0.41 &       11 &        0.27 &             0.37 &              0.02 &             0.76 &              0.04 \\
 Duck             & 204k  &        0.15 &        6 &        0.21 &              5.89 &        5 &        0.21 &              0.08 &       11 &        0.32 &           0.19 &      100 &        2.46 &           0.37 &       25 &        0.50 &             1.66 &              0.03 &             0.65 &              0.05 \\
 Mouse            & 214k  &        0.19 &        5 &        0.21 &              6.38 &        4 &        0.20 &              0.10 &       11 &        0.35 &           0.20 &      100 &        2.59 &           0.40 &       31 &        0.66 &             1.68 &              0.03 &             0.71 &              0.05 \\
 Wolf Skull       & 228k  &        0.23 &        5 &        0.26 &              8.01 &        4 &        0.29 &              0.13 &       13 &        0.57 &           0.26 &       44 &        1.50 &           0.51 &       32 &        0.94 &             0.38 &              0.02 &             0.91 &              0.04 \\
 Moses            & 258k  &        0.48 &        7 &        0.37 &              8.64 &        4 &        0.31 &              0.36 &       44 &        1.95 &           0.27 &      100 &        3.64 &           0.45 &      100 &        3.34 &             0.92 &              0.02 &             1.04 &              0.05 \\
 Rockerarm        & 271k  &        0.27 &        3 &        0.21 &              9.21 &        3 &        0.27 &              0.17 &        3 &        0.27 &           0.28 &       22 &        0.64 &           0.53 &       10 &        0.25 &             1.84 &              0.03 &             1.01 &              0.06 \\
 Pulley2          & 293k  &        0.24 &        4 &        0.29 &              8.81 &        4 &        0.29 &              0.14 &       10 &        0.49 &           0.28 &       96 &        3.49 &           0.69 &       23 &        0.72 &             3.07 &              0.04 &             1.07 &              0.07 \\
 Heraklion        & 350k  &        0.36 &       13 &        0.84 &             12.83 &        5 &        0.55 &              0.20 &       24 &        1.52 &           0.46 &      100 &        5.85 &           0.80 &      100 &        4.65 &             2.43 &              0.04 &             1.45 &              0.08 \\
 Julius Caesar    & 387k  &        0.30 &        7 &        0.41 &             11.61 &        9 &        0.63 &              0.16 &       12 &        0.68 &           0.36 &      100 &        4.89 &           0.75 &       36 &        1.44 &             5.22 &              0.06 &             1.28 &              0.09 \\
 Goyle            & 393k  &        0.32 &        2 &        0.20 &             11.39 &        2 &        0.26 &              0.21 &        4 &        0.32 &           0.37 &       14 &        0.67 &           0.79 &       11 &        0.43 &             8.43 &              0.07 &             1.32 &              0.10 \\
 Eros             & 476k  &        0.41 &        7 &        0.55 &             14.80 &        5 &        0.55 &              0.21 &       11 &        0.84 &           0.46 &      100 &        5.98 &           1.19 &       35 &        1.73 &             4.18 &              0.07 &             1.77 &              0.12 \\
 Roal             & 484k  &        0.47 &        4 &        0.47 &             16.50 &        4 &        0.59 &              0.38 &       12 &        1.20 &           0.53 &       64 &        4.40 &           1.08 &       19 &        1.10 &             1.31 &              0.04 &             2.08 &              0.09 \\
 Skeleton         & 494k  &        0.69 &        7 &        0.95 &             20.03 &        NaN &        NaN &              0.34 &       18 &        1.95 &           0.69 &      100 &        9.45 &           1.65 &       41 &        3.23 &             3.45 &              0.05 &             2.32 &              0.11 \\
 Bimba            & 502k  &        0.43 &        6 &        0.58 &             15.73 &        5 &        0.65 &              0.24 &       14 &        1.34 &           0.52 &      100 &        7.05 &           0.89 &       35 &        2.28 &             3.54 &              0.05 &             2.13 &              0.10 \\
 Oil Pump         & 570k  &        0.47 &       18 &        1.51 &             18.79 &       11 &        1.22 &              0.25 &       30 &        2.67 &           0.55 &       90 &        6.62 &           1.30 &       46 &        3.10 &             7.13 &              0.07 &             2.36 &              0.13 \\
 Antique Head     & 651k  &        0.53 &        4 &        0.51 &             20.10 &        4 &        0.60 &              0.28 &        9 &        0.84 &           0.64 &        7 &        0.60 &           1.37 &       44 &        2.82 &            14.65 &              0.11 &             2.42 &              0.18 \\
 Pulley           & 660k  &        0.54 &       18 &        1.70 &             21.62 &       14 &        1.66 &              0.28 &       27 &        2.77 &           0.62 &      100 &        8.47 &           1.48 &       46 &        3.43 &            12.80 &              0.10 &             2.68 &              0.16 \\
 Beard Man        & 691k  &        0.58 &        4 &        0.56 &             22.44 &        3 &        0.58 &              0.26 &        7 &        0.81 &           0.52 &        5 &        0.36 &           1.78 &        6 &        0.42 &            24.65 &              0.14 &             2.71 &              0.19 \\
 Red Circular Box & 701k  &        0.64 &        6 &        0.74 &             22.89 &        6 &        0.96 &              0.33 &       15 &        1.69 &           0.72 &      100 &        8.94 &           1.51 &       44 &        3.32 &            17.84 &              0.11 &             2.83 &              0.17 \\
 Dancing Children & 724k  &        0.69 &        7 &        0.89 &             23.09 &        9 &        1.35 &              0.40 &       21 &        2.54 &           0.68 &      100 &        9.27 &           1.97 &       58 &        4.98 &             6.34 &              0.09 &             2.77 &              0.17 \\
 John The Baptist & 750k  &        0.72 &       14 &        1.78 &             29.71 &        6 &        1.19 &              0.47 &       14 &        2.09 &           0.85 &      100 &       10.45 &           1.78 &       47 &        4.59 &             6.16 &              0.08 &             3.06 &              0.17 \\
 Ramses           & 826k  &        0.80 &        5 &        0.96 &             29.00 &        5 &        1.19 &              0.47 &       15 &        2.46 &           0.90 &      100 &       11.58 &           2.10 &       22 &        2.39 &             5.97 &              0.08 &             3.58 &              0.18 \\
 Nicolo Da Uzzano & 946k  &        0.77 &        7 &        1.02 &             30.04 &        5 &        1.02 &              0.46 &       12 &        1.66 &           0.85 &      100 &       12.03 &           2.01 &       33 &        3.24 &            19.50 &              0.16 &             3.65 &              0.25 \\
 Raptor           & 1m &        0.90 &       36 &        5.22 &             32.22 &        NaN &        NaN &              0.43 &       52 &        8.35 &           1.07 &      100 &       13.97 &           2.40 &       92 &       11.14 &             4.61 &              0.11 &             4.24 &              0.23 \\
 Nefertiti        & 1m &        0.89 &        4 &        0.94 &             34.34 &        4 &        1.18 &              0.57 &       14 &        2.79 &           1.09 &       57 &        8.41 &           2.55 &       14 &        1.87 &             9.23 &              0.11 &             4.56 &              0.22 \\
 Isidore Horse    & 1.1m &        0.94 &       10 &        1.65 &             35.42 &        5 &        1.28 &              0.50 &       24 &        3.98 &           1.10 &      100 &       14.46 &           2.47 &       74 &        8.89 &            23.25 &              0.17 &             4.54 &              0.28 \\
 Horse Head       & 1.3m &        2.36 &       20 &        5.71 &             53.36 &       11 &        4.80 &              1.00 &       74 &       19.51 &           2.18 &      100 &       26.10 &           4.43 &      100 &       22.69 &            12.88 &              0.14 &             6.55 &              0.31 \\
 Ram              & 1.3m &        2.72 &        3 &        2.18 &             57.16 &        NaN &        NaN &              1.21 &        5 &        2.53 &           2.32 &       83 &       21.68 &           4.68 &       25 &        5.49 &            17.59 &              0.15 &             7.05 &              0.33 \\
 Murex Romosus    & 1.8m &        2.30 &        5 &        2.56 &             72.19 &        4 &        2.87 &              1.01 &       26 &        8.99 &           2.32 &      100 &       29.59 &           5.16 &       24 &        6.30 &            38.34 &              0.25 &             9.06 &              0.43 \\
 XYZ Dragon       & 3.6m &        3.30 &        9 &        5.35 &            123.64 &        7 &        5.29 &              1.56 &       18 &       10.27 &           3.25 &      100 &       43.72 &           8.84 &       67 &       27.36 &            79.75 &              0.74 &            16.09 &              0.96 \\
\hline
\end{tabular}
}
\label{tab:comparisonall}
\end{table*}

\begin{table*}[h]
\caption{Comparison of our hierarchy construction and solver for \textbf{data smoothing} of a random function with smoothing coefficient $\alpha=$\num{1e-3} and tolerance of \num{1e-4} on non-manifold meshes and point clouds. The maximum number of iterations for iterative solvers is set to 100.}
\centering
\resizebox{\textwidth}{!}{
\begin{tabular}{@{}lr|rrr|rrr||rrr|rrr||rr|rr@{}}
\hline

 \multicolumn{1}{c}{\multirow{2}{*}{\textbf{Model}}} & \multicolumn{1}{c|}{\multirow{2}{*}{\textbf{\#Vert}}} & \multicolumn{3}{c|}{\textbf{\ourmethod~(Ours)}} & \multicolumn{3}{c||}{\textbf{\citet{Shi2006}}} & \multicolumn{3}{c|}{\textbf{AMG-RS}} & \multicolumn{3}{c||}{\textbf{AMG-SA}} & \multicolumn{2}{c|}{\textbf{Eigen}} & \multicolumn{2}{c}{\textbf{\textsc{Pardiso}}}  \\
\cline{3-18}
\multicolumn{1}{c}{} & \multicolumn{1}{c|}{} & \small{Hier} & \small{\#It} & \small{Solve} & \small{Hier} & \small{\#It} & \small{Solve} & \small{Hier} & \small{\#It} & \small{Solve} & \small{Hier} & \small{\#It} & \small{Solve} & \small{Fact.} & \small{Subst.} & \small{Fact.} & \small{Subst.}   \\
\hline
\multicolumn{18}{c}{\small{\textsc{non-manifold triangular meshes}}} \\
\hline
 Lakoon            & 188k  &        0.16 &        5 &        0.20 &              0.09 &       26 &        0.88 &           0.20 &      100 &        2.76 &           0.40 &       17 &        0.39 &             0.43 &              0.02 &             0.71 &              0.04 \\
 Indonesian Statue & 294k  &        0.26 &        7 &        0.42 &              0.17 &        7 &        0.44 &           0.30 &       37 &        1.45 &           0.63 &      100 &        3.53 &             0.90 &              0.03 &             1.18 &              0.06 \\
 Beethoven         & 383k  &        0.45 &        3 &        0.42 &              0.22 &        6 &        0.56 &           0.52 &        5 &        0.33 &           0.94 &       14 &        0.75 &             2.45 &              0.04 &             1.66 &              0.09 \\
 Bayon Lion        & 749k  &        1.44 &        4 &        1.20 &              0.71 &        8 &        1.67 &           1.37 &        7 &        1.07 &           2.45 &       10 &        1.28 &             6.26 &              0.08 &             3.75 &              0.18 \\
 Helmet Moustache  & 941k  &        2.04 &        5 &        2.07 &              0.74 &       22 &        4.97 &           2.10 &       15 &        2.97 &           3.44 &       17 &        2.79 &            24.99 &              0.14 &             5.56 &              0.26 \\
 Zeus              & 1.3m &        2.51 &        7 &        2.69 &              1.20 &       14 &        4.52 &           2.49 &       29 &        7.24 &           4.19 &      100 &       20.66 &            30.96 &              0.19 &             7.26 &              0.35 \\
 Alfred Jacquemart & 1.4m &        3.26 &        4 &        3.33 &              1.68 &       13 &        6.16 &           3.61 &        7 &        2.44 &           5.24 &       28 &        8.24 &             9.08 &              0.14 &             8.03 &              0.35 \\
\hline
\multicolumn{18}{c}{\small{\textsc{point clouds}}} \\
\hline
  Oil Pump      & 103k  &        0.07 &        4 &        0.07 &              0.04 &        6 &        0.11 &           0.10 &        7 &        0.09 &           0.19 &       12 &        0.13 &             0.17 &              0.01 &             0.30 &              0.02 \\
 Caesar Merged & 388k  &        0.29 &        4 &        0.30 &              0.17 &        7 &        0.52 &           0.41 &        7 &        0.39 &           0.83 &       31 &        1.40 &             4.81 &              0.06 &             1.50 &              0.10 \\
 Truck         & 1.2m &        0.96 &        6 &        1.39 &              0.68 &        9 &        2.14 &           1.27 &       12 &        2.30 &           3.69 &       46 &        6.88 &             5.81 &              0.15 &             5.24 &              0.29 \\
 Ignatius      & 1.4m &        1.25 &        6 &        1.67 &              0.78 &       15 &        4.14 &           1.58 &       30 &        6.44 &           4.43 &      100 &       17.64 &             8.89 &              0.18 &             6.11 &              0.35 \\
\hline
\end{tabular}
}

\label{tab:comparison_nm_pc}
\end{table*}

\end{document}